\documentclass{elsart}
\usepackage{psfig}
\usepackage{endfloat}

\newcommand{\cal}{\bf}
\newcommand{\be}{\begin{equation}}
\newcommand{\ee}{\end{equation}}
\newcommand{\bea}{\begin{eqnarray}}
\newcommand{\eea}{\end{eqnarray}}
\newcommand{\gbf}[1]{\mbox{\boldmath ${#1}$}} 

\newcommand{\eg}{{\em e.g. }}

\def \Re{{\rm I\kern -1.6pt{\rm R}}}
\def \Expect{{\rm I\kern -1.6pt{\rm E}}}

\journal{Physica D}
\begin{document}

\begin{frontmatter}
\title{Consistent nonlinear dynamics: identifying model inadequacy}

\author[MATHS,ENG,LSE]{Patrick E. McSharry\corauthref{cor}}
\corauth[cor]{Corresponding author.} \ead{mcsharry@maths.ox.ac.uk}
\ead[url]{http://www.cats.lse.ac.uk}
\author[MATHS,LSE]{Leonard A. Smith}

\address[MATHS]{Mathematical Institute, University of Oxford,
Oxford OX1 3LB, UK.}
\address[ENG]{Department of Engineering Science, University of Oxford,
Oxford OX1 3PJ, UK.}
\address[LSE]{Centre for the Analysis of Time Series,
London School of Economics, London WC2A 2AE.}

\begin{abstract}
Empirical modelling often aims for the simplest model consistent
with the data. A new technique is presented which quantifies the
consistency of the model dynamics as a function of location in
state space. As is well-known, traditional statistics of nonlinear
models like root-mean-square (RMS) forecast error can prove
misleading. Testing consistency is shown to overcome some of the
deficiencies of RMS error, both within the perfect model scenario
and when applied to data from several physical systems using 
previously published models. In particular, testing for consistent
nonlinear dynamics provides insight towards (i) identifying when a
delay reconstruction fails to be an embedding, (ii) allowing state
dependent model selection and (iii) optimising local neighbourhood
size. It also provides a more relevant (state dependent) threshold
for identifying false nearest neighbours.
\end{abstract}

\end{frontmatter}

\section{Introduction}
The construction of mathematical models whose dynamics
reflect the observations of physical systems is arguably the
fundamental task of physics \cite{feynman}. A model's ability to
reproduce the observed dynamics is often quantified through the
distribution of forecast errors. Indeed in traditional time series
analysis the ``optimal model'' is defined as that which minimises
the root-mean-square (RMS) error \cite{chatfield89}. For a
nonlinear system, this approach has the undesirable property that
it may reject the model which generated the data. In this paper, a
new test is introduced which quantifies the level of consistency
between each iteration of the model and the observations; this
test is of particular value for nonlinear models where uncertainty
in the data due to observational noise limits the utility of
statistics like RMS error
\cite{chatfield89,mcsharry99a,smith97fermi}. Ideally, a model will
admit trajectories which shadow the entire dataset to within the
limits set by observational noise
\cite{smith97fermi,smith00,grebogi90}; testing for consistent
nonlinear dynamics (CND) provides a simple, computationally
tractable, necessary condition for shadowing. The CND approach is
applicable to models based on first principles
\cite{palmer00,orrell01} and data-based empirical models
\cite{kantzbook,farmers87,priestly81,brooml88,casdagli89,smith92}.
The aim is to confirm the simplest model consistent with the
data, but none simpler.

As a test of the dynamics, the CND approach quantifies consistency
as a function of location in model-state space. It is novel in
that it (i) identifies regions where the model is notably
imperfect, (ii) can suggest regions where an embedding fails (that
is, where no deterministic model is consistent), (iii) provides a
rational for setting a (local) threshold for the method of false
nearest neighbours \cite{kennel92}, and (iv) provides a means of
selecting and weighting models for multi-model ensemble prediction
\cite{smith97fermi,palmer00}. Models may be either linear or
nonlinear, and the results below are easily generalised to
stochastic models \cite{chatfield89,tong90}, although the
discussion here is restricted to the case of deterministic models
and additive observational noise. While most of the examples below
evaluate data-based empirical models, the consistent nonlinear
dynamics approach is applicable to {\it any} dynamical model: it
evaluates the particular model along with background modelling
assumptions such as stationarity and the assumed form of
observational noise.

Basic goals and methodology are presented in Section
\ref{s:method}. Section 3 provides a brief and selective review of
data-based model building in delay space. Two mathematical systems
are then used to illustrate the method in section
\ref{s:theoretical} and applications to previously published
models of three widely studied physical systems, each thought to
be chaotic, is given in section \ref{s:physical}. The implications
of these results for nonlinear modelling are discussed and
interpreted in section \ref{s:discussion} and conclusions are
given in section \ref{s:conclusion}.

\begin{figure}
\centerline{\psfig{file=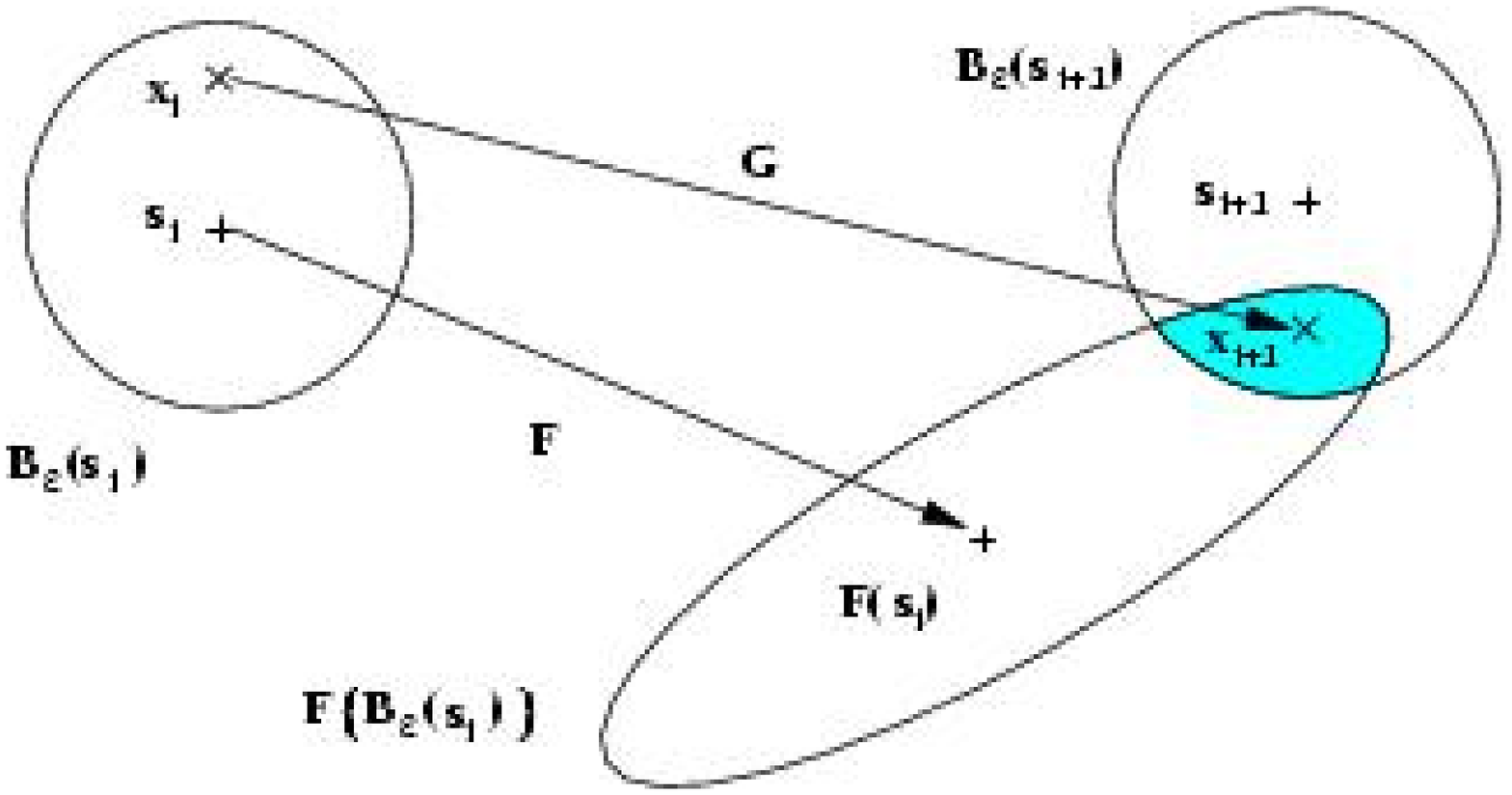,width=\linewidth}}
\caption{Consistent prediction: a trajectory of ${\bf G}$,
with initial and final positions ${\bf x}$ ($\times$), corresponding
observed positions ${\bf s}$ (+), their consistency balls,
${\cal B}_{\epsilon}({\bf s}_i)$ and
${\cal B}_{\epsilon}({\bf s}_{i+1})$  ({\it circles}), and
images of model-states which are consistent with ${\bf s}_i$,
${\bf F} \{ {\cal B}_{\epsilon}({\bf s}_i)\}$ ({\it ellipse}).
The intersection of ${\bf F} \{ {\cal B}_{\epsilon}({\bf s}_i)\}$ with
${\cal B}_{\epsilon}({\bf s}_{i+1})$ signifies consistency.}
\label{f:conmap}
\end{figure}

\section{Methodology}
\label{s:method}

In this section, a constraint on state-dependent forecast error is
derived; this forms the basic consistency test of CND. A number of
simplifying assumptions are made to ease the derivation, often
these can be relaxed and more general solutions deployed
numerically (at the cost of simplicity and computational time).

The most common means for quantifying the quality of a model are
based on the statistics of prediction errors \cite{chatfield89}
(for alternatives, see
\cite{mcsharry99a,kostelich90,judds04,beven01}). Consider a model
${\bf F}$, with observations ${\bf s}_i \in \Re^{m_s}$
corresponding to system states ${\bf x}_i \in \Re^{m_x}$ of the
`true' deterministic system ${\bf G}$. Assume $m_s = m_x = m$ in
the following\footnote{Note that in general the model-state space
will differ from the true state space of the system (assuming one
exists); the projection from one space to another poses several
foundational difficulties. This projection operator will be taken
to be the identity here (for discussion see
\cite{smith97fermi,beven02} and references therein).}. 
Also assume additive measurement errors 
$\gbf{\eta}_i$ (i.e. ${\bf s}_i = {\bf x}_i + \gbf{\eta}_i$). 
In this case, the one-step prediction error
may be decomposed as \bea
{\bf E}_{\rm pred} &=& {\bf s}_{i+1} - {\bf F}({\bf s}_i) \nonumber \\
&=&  ({\bf x}_{i+1} + \gbf{\eta}_{i+1}) - {\bf F}({\bf x}_i + \gbf{\eta}_i)
\nonumber \\
&=& \left[{\bf G}({\bf x}_i) - {\bf F}({\bf x}_i)\right]
+ \left[\gbf{\eta}_{i+1} + {\bf F}({\bf x}_i)
- {\bf F}({\bf x}_i + \gbf{\eta}_i) \right] \nonumber \\
&=& {\bf E}_{\rm model} + {\bf E}_{\rm noise}, \label{e:pedecomp}
\eea where ${\bf E}_{\rm model}$ represents model
inadequacy\footnote{A given model may be inconsistent due to {\it
parametric error} (inaccurate parameter values) or due to {\it
structural error} (the model class from which the particular model
equations are drawn does not include a process that might have
generated the data, given the observational noise\cite{judds04}).
In the second case the model equations themselves are inadequate.} 
and ${\bf E}_{\rm noise}$ represents error due to observational uncertainty.
Viewing the prediction error ${\bf E}_{\rm pred}$ as a sum of
these two distinct sources of error highlights three facts: (i)
prediction errors may be due to noise alone, and need not arise
from model error, (ii) model error is obscured by observational
uncertainty, and can only be accurately assessed if this
uncertainty is taken into account, and (iii) prediction errors may
manifest extreme fluctuations throughout model-state space due to
variations in model sensitivity. Note the similarity of this
decomposition with that for ``model drift" used in operational
weather forecasting \cite{orrell01,orrell}.

\begin{figure}
\centerline{\psfig{file=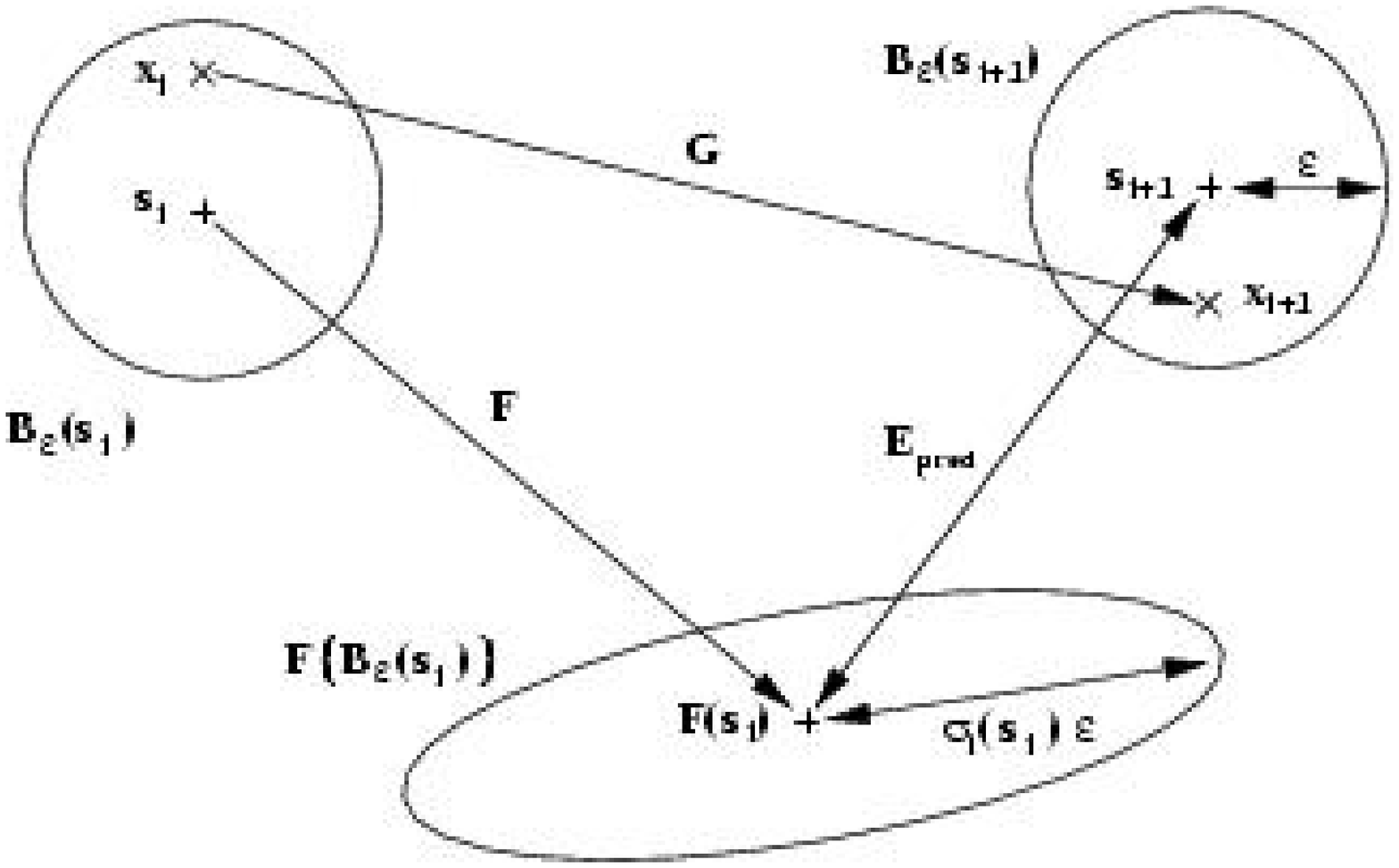,width=\linewidth}}
\caption{Inconsistent prediction: as Fig. 1. In this case
no model trajectory exists which
is consistent with the observational uncertainty since
${\bf F} \{ {\cal B}_{\epsilon}({\bf s}_i)\}$ does not intersect
${\cal B}_{\epsilon}({\bf s}_{i+1})$.
}
\label{f:incmap}
\end{figure}

Prediction error due to additive noise in a perfect model may be
expressed as a Taylor series: 
\be {\bf E}_{\rm noise} =
\gbf{\eta}_{i+1} - {\bf J}({\bf x}_i) \gbf{\eta}_i - \frac{1}{2}
\gbf{\eta}^T_i {\bf H}({\bf x}_i) \gbf{\eta}_i - \cdots
\label{e:enoisete} 
\ee 
where ${\bf J}({\bf x})$ and ${\bf H}({\bf
x})$ are the Jacobian and Hessian of ${\bf F}({\bf x})$. The
distribution of ${\bf E}_{\rm noise}$ in (\ref{e:enoisete})
reflects how variations in the local derivatives of a particular
model structure affects the local prediction error. When the
distribution of $\gbf{\eta}_i$ is known, the distribution of ${\bf
E}_{\rm noise}$ can be determined from (\ref{e:enoisete}) and,
since ${\bf E}_{\rm pred}$ is known, regions of model-state space
with large ${\bf E}_{\rm model}$ can be identified. Figures
\ref{f:conmap} and \ref{f:incmap} illustrate this in model-state
space where, for simplicity, the initial uncertainty is bounded
(constrained to be within the circle) and the dynamics are locally
linear (the circle evolves into an ellipse). The system state at
initial time ${\bf x}_i$ and at final time ${\bf x}_{i+1}$ are
marked by {\it crosses}. For model-state vectors ${\bf s}_i \in
\Re^m$ and observational uncertainty uniformly distributed in a
sphere of radius $\epsilon$, there exists an associated ball of
consistent model-states \mbox{${\cal B}_{\epsilon}({\bf s}_i) =
\left\{\gbf{\xi}: \left|\left| \gbf{\xi} - {\bf s}_i
\right|\right| \leq \epsilon \right\}$}. In practice, only the
observed positions ${\bf s}_i$ ({\it pluses}) and their associated
consistency balls ${\cal B}_{\epsilon}({\bf s}_i)$ ({\it circles})
are known. The consistency of a model's forecast may be tested by
evolving the ball ${\cal B}_{\epsilon}({\bf s}_i)$ to the final
time. This set of evolved states, ${\bf F} \{ {\cal
B}_{\epsilon}({\bf s}_i)\}$, will resemble an ellipsoid if the
evolution is locally linear (i.e. when $\epsilon$ is sufficiently
small and ${\bf F}$ is sufficiently smooth). If ${\bf F} \{ {\cal
B}_{\epsilon}({\bf s}_i)\}$ and ${\cal B}_{\epsilon}({\bf
s}_{i+1})$ intersect, then the prediction is internally consistent
(Fig. \ref{f:conmap}), otherwise it is deemed inconsistent (Fig.
\ref{f:incmap}). The overlap of ${\bf F} \{ {\cal
B}_{\epsilon}({\bf s}_i)\}$ and ${\cal B}_{\epsilon}({\bf
s}_{i+1})$ implies the existence of a model trajectory which lies
inside the ball ${\cal B}_{\epsilon}({\bf s}_i)$ at the initial
time and also inside the ball ${\cal B}_{\epsilon}({\bf s}_{i+1})$
at the final time; that trajectory is consistent with both the
model dynamics and the observations.

The most pedestrian version of CND is simply to verify this
necessary condition. The idea is simply to check whether the
largest observational error in the worst direction is likely to
account for the observations. In practice, obtaining ${\bf F} \{
{\cal B}_{\epsilon}({\bf s}_i)\}$ by propagating a ball of
model-states in $\Re^m$ may prove to be computationally expensive.
A first order approximation provides a linear consistency check
which is easy to implement. The linear approximation of
\mbox{(\ref{e:enoisete}) is} \be {\bf E}_{\rm noise} =
\gbf{\eta}_{i+1} - {\bf J}({\bf x}_i) \gbf{\eta}_i.
\label{e:enoisete1} \ee The first singular value, $\sigma_1({\bf
s}_i)$, of the Jacobian \cite{smith97fermi,palmer00} describes the
largest possible (linear) magnification which corresponds to the major
axis of \mbox{${\bf F} \{ {\cal B}_{\epsilon}({\bf s}_i) \}$}.
Note from Fig. \ref{f:incmap} that there is no initial condition
consistent under the model if the prediction error is greater than
the magnitude of the major axis of \mbox{${\bf F} \{ {\cal
B}_{\epsilon}({\bf s}_i) \}$} plus the radius of ${\cal
B}_{\epsilon}({\bf s}_{i+1})$. If the noise level is large
relative to the local curvature of the (model's) solution
manifold, then the linear approximation may be generalised for
locally nonlinear models; for local linear models this result is
exact.

The expected value of the observational noise divided by the
(local) length scale at which quadratic terms become important
provides a fundamental ratio for each local linear prediction; if
this ratio is large one must consider local quadratic models. In
regions where this ratio is small, and the length scale at which
quadratic terms become important is also small relative to near
neighbour distances, it is advantageous to take smaller
neighbourhoods thereby improving the local linear model. CND can
identify such regions.

Note that the details of the consistency test depends on the type
of state space being employed. The state space may be constructed
using either multivariate data or a delay reconstruction of
univariate data. In the first case, each component of the
model-state vector must be forecast and evaluated while in the
case of the delay reconstruction only one component need be
forecast, the others generated by a shift operator; this
difference will have an impact on tests of CND. 
The type of uncertainty in the
measurements may also be either bounded (\eg truncation error) or
unbounded (\eg additive Gaussian noise). In the former, the initial
observational uncertainty resulting from bounded measurement
errors lies in a hyper-cube (that is, a box); given Gaussian
noise, an isopleth of equal probability in the space will be a
sphere (or an ellipse, depending on the metric adopted). The
discussion below is framed in terms of quantization error, but
this can easily be interpreted in terms of, say. the 99.9\%
isopleth of an unbounded noise distribution.

Many of the applications presented below assume a delay
reconstructed state space with truncation errors in the
observations, implying that $\gbf{\eta}$ is uniformly distributed
inside a hyper-cube. The diagonal of any $m$-dimensional hyper-cube of 
side $\epsilon$ may be stretched to at most 
$\sigma_1({\bf s}_i) \sqrt{m} \epsilon$  
and uncertainty in the (scalar) verification implies an additional
factor of $\epsilon$, so that the consistency measure is \be {\cal
C}({\bf s}_i,\epsilon) = \frac{|| {\bf E}_{\rm pred} ||}
{\left[\sigma_1({\bf s}_i) \sqrt{m} + 1\right]  \epsilon}.
\label{e:con} \ee This definition of linear consistency explicitly
states its dependence on both the observational uncertainty,
quantified by $\epsilon$, and the local stretching due to the
model structure, expressed through $\sigma_1({\bf s}_i)$. Since
the value of $\sigma_1({\bf s}_i)$ is obtained from the model,
${\cal C}({\bf s}_i,\epsilon) \leq 1$ is a necessary but not
sufficient condition. If the model's $\sigma_1({\bf s}_i)$ is too
small the inconsistency will be detected, but {\it not} if
$\sigma_1({\bf s}_i)$ is too large. This approach can be
generalised to explicitly test for overlap between the two
regions, and can be extended beyond one-step forecasts
\cite{mcsharry99a}; nevertheless the simplest case is used here
to more clearly illustrate the procedure.

Note that in some regions of model-state space, large prediction
errors are expected due to large values of $\sigma_1({\bf s}_i)$.
Alternatively, even relatively small prediction errors may be
inconsistent in a region of state space where $\sigma_1({\bf
s}_i)$ is small. Lone inconsistent predictions may be due to
outliers \cite{chatfield89} in the data stream, so significance is
afforded by a localised accumulation of inconsistent predictions.
Such accumulations may be due to failure of the delay space to
provide an embedding, erroneous parameter values, or structural
error in the model. In a nonlinear dynamical system these are
intrinsically mixed; and they are indistinguishable given only a
single model (see \cite{schroer98,dphil99}). Of course, regions of
model error can easily be located and interpreted when both the
model and system are known analytically. While this perfect model
scenario (PMS) is used in section \ref{s:theoretical} to
illustrate the technique, much interest lies in physical systems
where no perfect model (or model class) is known. Physical systems
are considered in section \ref{s:physical}, where the CND approach
is applied to three widely studied physical systems using
previously published nonlinear models.   First, the construction
of models in delay space is briefly reviewed.

\section{Empirical models}
\label{s:empirical}

This section provides a summary of the  data-driven modelling
techniques which are used below. These models provide a mapping
$F: \Re^m \to \Re$, ${\bf s}_i \mapsto {\hat s}_{i+\tau_p}$ from
the present observed state vector ${\bf s}_i$ to an estimate
${\hat s}_{i+\tau_p}$ of the future observed value $s_{i+\tau_p}$
where $\tau_p$ is the prediction horizon. In the discussion in
section \ref{s:method} the model state space was taken to include
simultaneous observations of the physical variables. In the case
of univariate observations, ${\bf s}_i$ will be represented by a
$m$-dimensional delay reconstruction \cite{takens81,sauer91} \be
{\bf s}_i = \left[ s_{i-(m-1)\tau_d},\ldots, s_{i-\tau_d}, s_i
\right] \label{e:delrec} \ee where $\tau_d$ is the time delay. As
noted above, the use of a delay reconstruction allows CND to focus
only on the last component of the state vector.

Local polynomials can be used to provide an approximation of the
nonlinear dynamics in a neighbourhood ${\cal B}(s_i)$ about a
reconstructed state vector ${\bf s}_i$. Two local polynomial model
formulations  are employed below: 
\be F({\bf s}_i) = a_0 +
\sum_{j=1}^m a_j s_{i-(j-1)\tau_d}, 
\label{e:loclin} 
\ee 
for a local
linear model, and 
\be 
F({\bf s}_i) = a_0 + \sum_{j=1}^m a_j s_{i-(j-1)\tau_d} + 
\sum_{j=1}^m \sum_{l=j}^m a_{jm+l} s_{i-(j-1)\tau_d}
s_{i-(l-1)\tau_d}, 
\label{e:locquad} 
\ee 
for a local quadratic model. The
model parameters or polynomial coefficients ${\bf a}$ are
determined by solving the over-determined system of linear
equations formed by substituting the $k$ pairs $\{ {\bf
s}_{\kappa(l)},s_{\kappa(l)+\tau_p} \}_{l=1}^k$ in (\ref{e:loclin}) or
(\ref{e:locquad}) where $\kappa(l)$ gives the indices of the
points found in the local neighbourhood ${\cal B}({\bf s}_i)$,
specified either by fixing the number of neighbours $k$ or by
choosing a radius $r$. Both local linear (LL)
\cite{farmers87,priestly81} and local quadratic (LQ)
\cite{casdagli89,smith92} are employed for making predictions in this
paper.

Radial Basis Function (RBF) models of low dimensional chaotic
systems (see \cite{brooml88,casdagli89,smith92} and references
thereof) provide a global, empirically based, nonlinear system of
equations. An RBF model is of the form \be F({\bf s}_i) =
\sum_{j=1}^{N_c} a_j \phi(|| {\bf s}_i - \gbf{\xi}_j ||) + L({\bf
s}_i), \label{e:rbf} \ee where $\phi(r)$ is the radial basis
function, $\gbf{\zeta}_j$ are centres, ${\bf a} =
\{a_j\}_{j=1}^{N_c}$ contains the model parameters and $L$
represents an additional global linear (or higher order)
polynomial which is often found to improve the approximation
\cite{casdagli89,smith92}. The two basis functions used in this
paper are cubic $\phi(r) = r^3$ and Gaussian $\phi(r) = e^{-r^2/2
\sigma^2}$.  Following \cite{smith92} the centres were chosen
uniformly in the reconstructed state space in regions where the
data existed and $\sigma$ was given by the average Euclidean
distance between centres which were nearest neighbours.

An attractive feature of the RBF model is that its parameters can
be determined via a single singular value decomposition
\cite{press92}. The model parameters ${\bf a} = \{a_j\}_{j=1}^k$
are determined by solving the linear system of equations
\mbox{${\bf b} = \gbf{\Phi} {\bf a}$}, where the design matrix is
\mbox{$\Phi_{ji} = \phi(||{\bf s}_i -\gbf{\zeta}_j||)$}, the
images are represented as ${\bf b} = \{ s_{i+\tau_p}
\}_{i=1}^{N_l}$ and $N_l$ is the size of the learning data set.
This is achieved by obtaining parameters ${\bf a}$ which minimise
$\chi^2 = || {\bf b} - \gbf{\Phi} {\bf a} ||^2$. Both $\chi^2$ and
$||{\bf a}||$ are minimised by choosing \mbox{${\bf a} =
\gbf{\Phi}^\dag {\bf b}$}, where $\gbf{\Phi}^\dag$ is the
Moore-Penrose pseudo-inverse of $\gbf{\Phi}$ \cite{press92}.
Alternatives to using the default weighting (which is, roughly,
uniform with respect to the invariant measure) are discussed in
Section \ref{sec:annulus} below (see also Smith \cite{smith92}).
In both cases, the solution is linear in the parameters.

\section{Forecasting where a perfect model exists}
\label{s:theoretical}

In this section models with known inadequacy are used to
illustrate that, at least within the perfect model scenario, CND
correctly diagnoses model inadequacy. Two well studied
mathematical dynamical systems are used. When forecasting physical
systems, of course, there is no reason to believe that a perfect
model exists. In such cases, methods like CND must be evaluated
based on their ability to provide useful information. This is
illustrated for a variety of physical systems in section
\ref{s:physical}. In both cases, it is crucial to distinguish the
model(s) from the system that generated the data (whether the
system be a set of equations or a measuring device).

\subsection{Ikeda map}
In the late 1970's, Ikeda \cite{ikeda79,ikeda80} pointed out that
the plane-wave model of a bistable ring cavity exhibited period
doubling cascades to chaos. Hammel {\it et al.} \cite{hammel85}
then extracted a complex difference equation relating the field
amplitude at the $(i+1)^{st}$ cavity pass to that of a round trip
earlier.  The amplitude $x$ and phase $y$, corresponding to the
real and imaginary parts of the field are related by \bea
x_{i+1} &=& 1 + \mu(x_i \cos \theta - y_i \sin \theta), \nonumber \\
y_{i+1} &=& \mu(x_i \sin \theta + y_i \cos \theta), \nonumber \\
\theta &=& a - b/(x_i^2 + y_i^2 + 1). \label{e:ikeda} \eea This
map is chaotic for parameter values $a = 0.4, b = 6.0$ and $\mu =
0.9$ \cite{hammel85}.

A family of imperfect models of this system was developed by
Judd and Smith \cite{judds04} by considering the Taylor expansions
of the sinusoidal functions in (\ref{e:ikeda}). The translation
$\theta = - \pi + \omega$ is employed \cite{judds04} to yield an
expansion about $-\pi$ which is near the middle of the range of
$\theta$. Neglecting terms of sixth-order and higher
yields the model equations: \bea \cos \theta &=& cos(-\pi +
\omega) = -1 + \omega^2 - \omega^4/24
\\ \nonumber \sin \theta &=& sin(-\pi + \omega) = -\omega +
\omega^3/6 - \omega^5/120. \label{e:ikedatrun} \eea

\begin{figure}
\centerline{\psfig{file=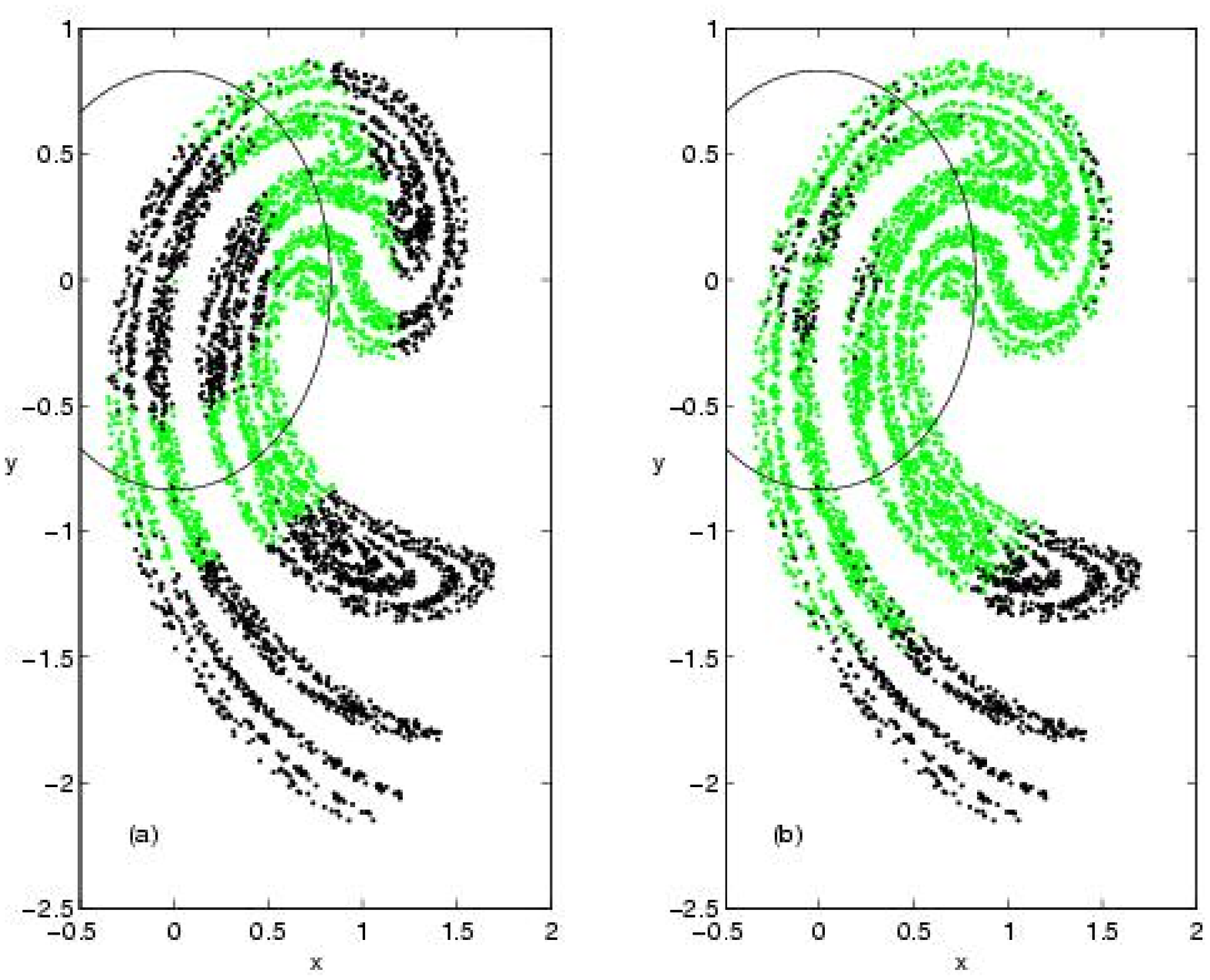,width=\linewidth}}
\caption{Consistent (grey) and inconsistent (black) predictions of the
full state space Ikeda map using (a) third order truncation
and (b) fifth order truncation of (\ref{e:ikedatrun}). The black line
indicates where the truncated models are exact ($\theta = -\pi$).}
\label{f:ikedacon}
\end{figure}

Adopting the Ikeda map (\ref{e:ikeda}) as the system, time series
were generated for both the $x$ and $y$ variables with additive
noise. The  measurement errors were uniformly distribution on
$[-\epsilon,\epsilon]$ with $\epsilon = 0.01$. Both a third-order
truncated model and a fifth-order truncated model were used to
make one step ahead predictions of the noisy bivariate time
series. These predictions were coloured grey when consistent and
black when inconsistent (Fig. \ref{f:ikedacon}). The circle $x^2 +
y^2 = \frac{b}{a+\pi}-1$, corresponding to $\theta = - \pi$ and
$\omega = 0$ (where the truncated models are exact) is also shown.
As expected the models provide consistent predictions in the
vicinity of a band around the circle and the width of this band
increases with the addition of higher order terms in
(\ref{e:ikedatrun}).

It is often the case that no sufficiently relevant model equations
are available and that an empirical, data-driven approach is
required. There are two common strategies in this case: to
construct a local model as each prediction is required or to
construct a single, global model. For clarity, only local linear
models and local quadratic models are employed below, while the
empirical global models used are based on radial basis functions;
both approaches were introduced briefly in section
\ref{s:empirical}.

\begin{figure}
\centerline{\psfig{file=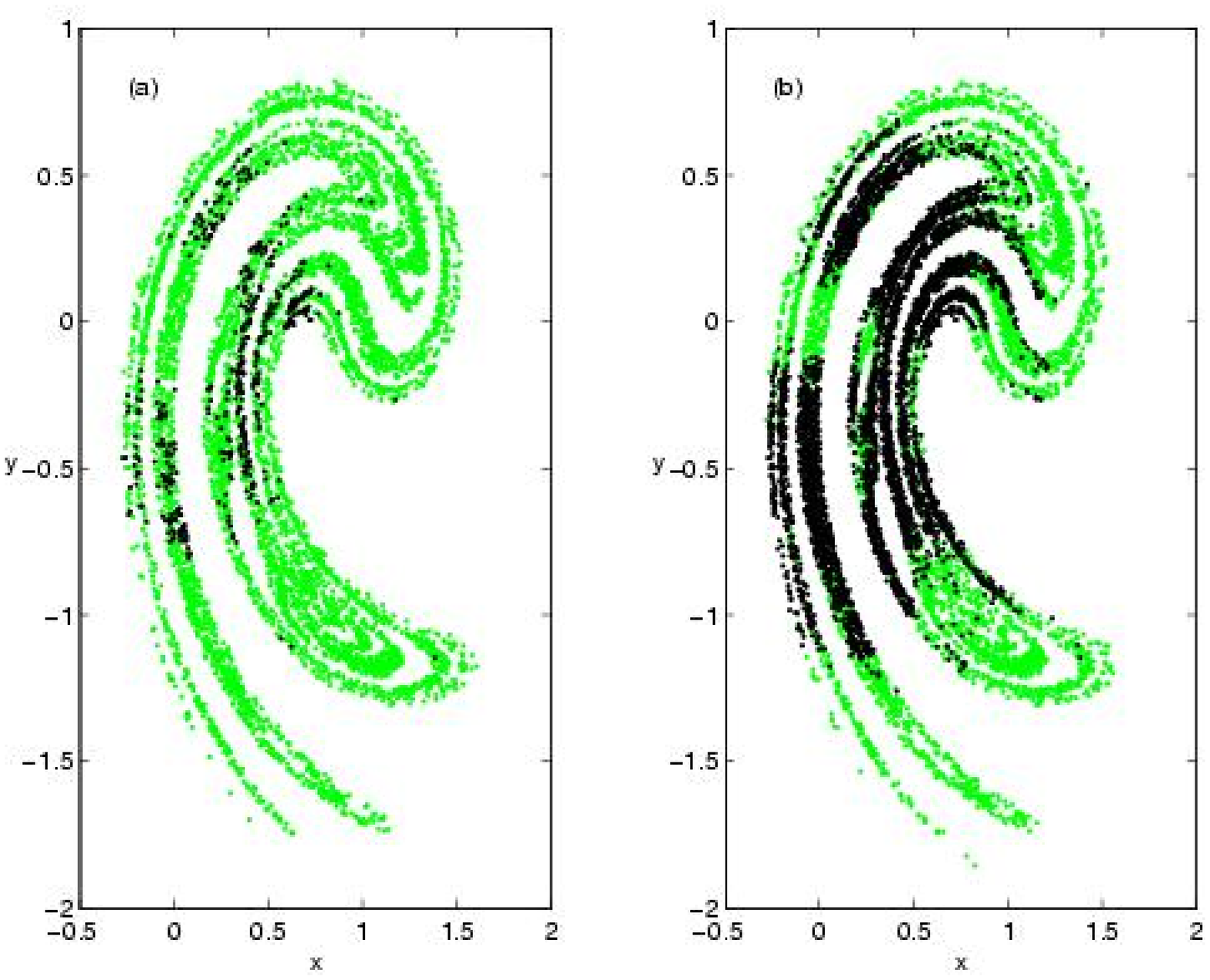,width=\linewidth}}
\caption{Consistent (grey) and inconsistent (black) predictions of the
full state space Ikeda map using local linear prediction with neighbourhood
radii of (a) 0.15 and (b) 0.20.}
\label{f:ikedallnm2r}
\end{figure}

CND considers the local neighbourhood around the base point of the
prediction. In this example, a linear model is fit to describe the evolution
of historical base points within this neighbourhood towards their
future images. This linear model is then applied to the base point
to yield a prediction of its future image. The optimal size of the
local neighbourhood ${\cal B}({\bf x},r)$, at a point ${\bf x}$
with radius $r$, depends on (i) the data density in ${\cal B}({\bf
x},r)$, (ii) the magnitude of the neglected nonlinear terms in
${\cal B}({\bf x},r)$ and (iii) the measurement uncertainty of
points in ${\cal B}({\bf x},r)$ (for a discussion, see
\cite{smith94c,bollt00}).  Local linear predictions of the noisy
measurements of both $x$ and $y$ variables of the Ikeda map using
$r = 0.15$ and $r = 0.2$ are illustrated in Fig.
\ref{f:ikedallnm2r}. The fraction of inconsistent predictions were
$f_{inc} = 0.025$ and $f_{inc} = 0.27$ for $r = 0.15$ and $r =
0.2$ respectively. The predictor with the smaller radius of $r =
0.15$ produces fewer inconsistent predictions because the local
linear model is better able to capture the dynamics in this
neighbourhood. This makes sense, in that for larger values of $r$
there are more locations where the local quadratic terms play a
significant role and thus the linear model (which is blind to
these effects) fails to yield consistent predictions.

\begin{figure}
\centerline{\psfig{file=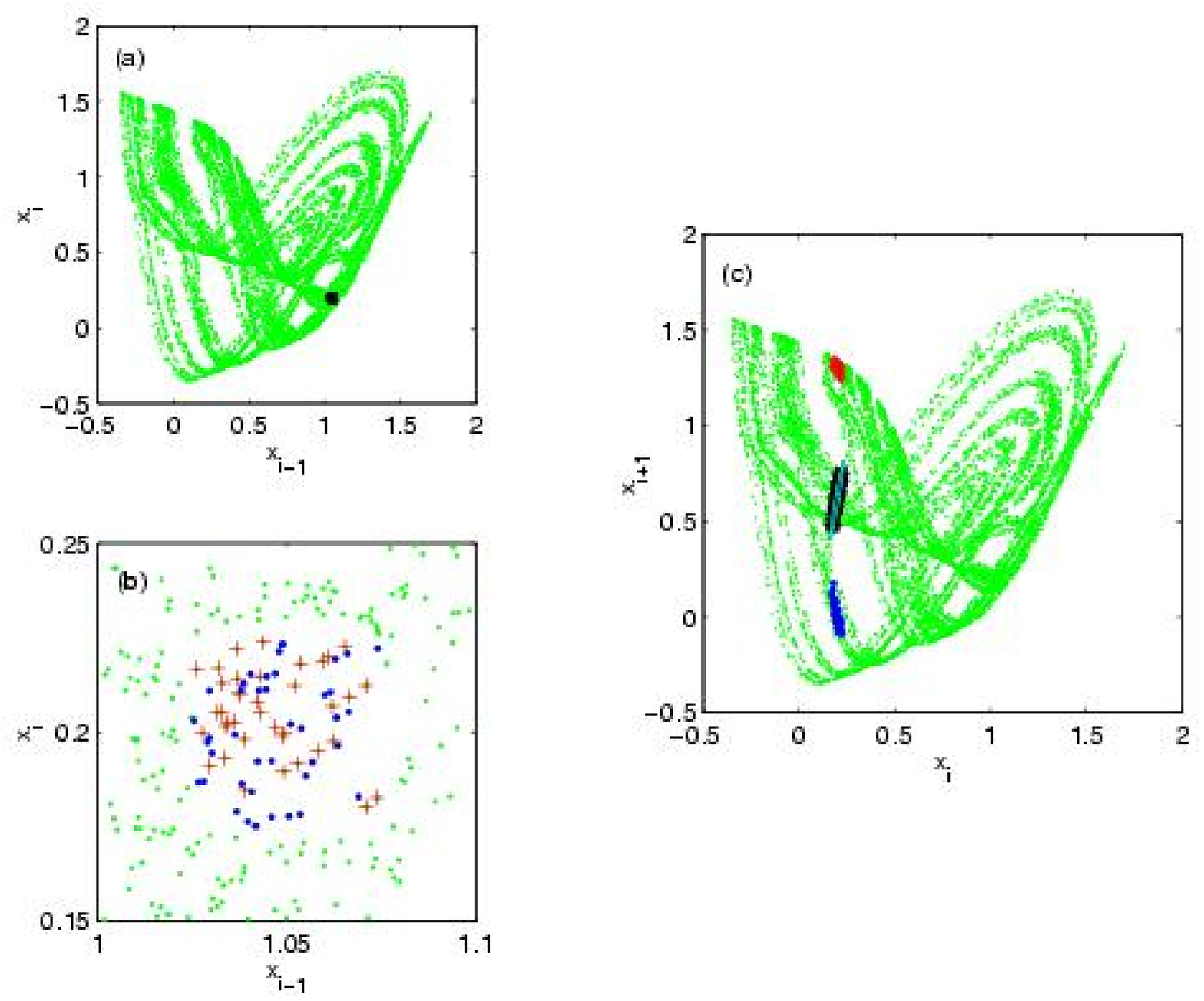,width=\linewidth}}
\caption{An illustration of self-intersection in a $m = 2$
reconstruction of the Ikeda map: (a) the reconstructed state
space $(x_{i-1},x_i)$ showing observed points within a square centred at
the base-point (1.05,0.2) being consistent with measurement errors of
magnitude $\epsilon = 0.025$,
(b) a zoom-in showing pre-images of points which have large $x_{i+1}$ (pluses)
and points which have small values of $x_{i+1}$ (dots) and (c)
the reconstructed state space $(x_i,x_{i+1})$ with images from (b),
corresponding local linear predictions (black) and consistency ellipse (grey).}
\label{f:ik2dintsv}
\end{figure}

The consistency analysis can also be used to explore whether or not
there are self-intersections within a reconstructed state space.
For the Ikeda map, there are many self intersections using a reconstruction
dimension of $m = 2$ (Fig \ref{f:ik2dintsv}a).
These self-intersections imply that nearby points, often called
false neighbours, (Fig. \ref{f:ik2dintsv}b) will
have images in different parts of the attractor (Fig \ref{f:ik2dintsv}c).
The singular value decomposition of the local linear map provides a means of
calculating the expected region (an ellipse under the linear approximation)
where the images should fall. In this case of self-intersections, the
consistency analysis correctly shows that the predictions should lie
between the two extremes of the images, coloured red and blue
(Fig. \ref{f:ik2dintsv}c) and
the resulting inconsistent predictions reflect
a failure of this reconstruction due to $m$ not being sufficiently large.

\begin{figure}
\centerline{\psfig{file=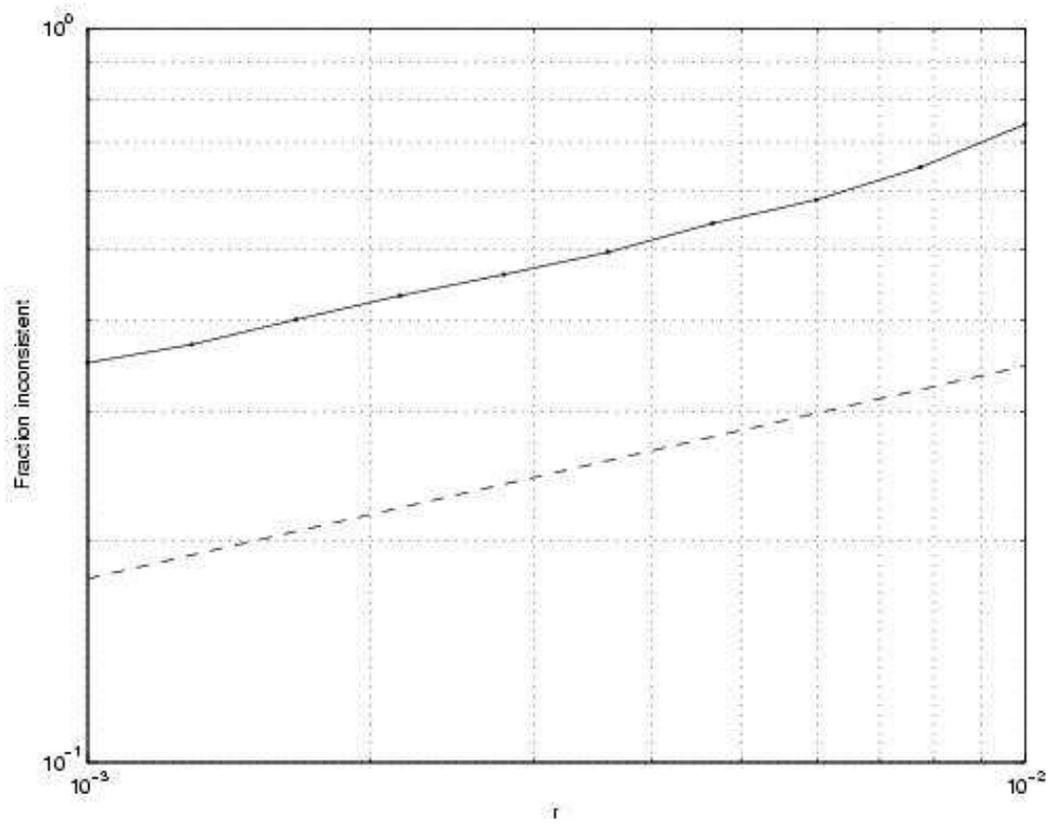,width=\linewidth}}
\caption{Fraction of inconsistent predictions as a function of
the neighbourhood radius $r$ used for fitting the local linear
model (solid). The dashed line indicates the theoretical slope given
by $m - D_1 = 2 - 1.71 = 0.29$.}
\label{f:ikedaconr}
\end{figure}

Schroer {\it et al.} \cite{schroer98} have shown that the fraction
of the attractor which yields poor predictions due to
self-intersections scales as $r^{m - D_1}$ where $r$ is the size
of the neighbourhood used for constructing the local constant
predictor and $D_1$ is the information dimension. A noise-free
time series of the $x$-coordinate of the Ikeda map (\ref{e:ikeda})
significant to four decimal places was generated and used to provide
a reconstructed state space with $m = 2$. Local linear predictions
with a neighbourhood of radius $r$ were analysed for consistency
against a noise threshold of $\epsilon = 5 \times 10^{-5}$ using
(\ref{e:con}). The fraction of inconsistent points $f_{inc}$
decreases with the size of the local neighbourhood radius $r$
according to $f_{inc} \propto r^{m-D_1}$ (Fig. \ref{f:ikedaconr}),
in agreement with the theoretical calculations \cite{schroer98}.
$D_1$ was estimated using the Kaplan-Yorke conjecture
\cite{kaplany79} \be D_1 = D_{Lyap} \equiv k + \frac{\sum_{i=1}^k
\lambda_i}{\lambda_{k+1}} \ee where $D_{Lyap}$ is the Lyapunov
dimension, $\lambda_i$ are the Lyapunov exponents and $k$ is
defined such that $\sum_{i=1}^k \lambda_i \geq 0$ and
$\sum_{i=1}^{k+1} \lambda_i < 0$.  For the Ikeda system with
parameter values $a = 0.4, b = 6.0$ and  $\mu = 0.9$, $D_{Lyap} =
1.71$ (for discussion, see McSharry \cite{dphil99}).

\subsection{False nearest neighbours}

The method of false nearest neighbours introduced by Kennel {\it et al.}  
(see \cite{kennel92} and references thereof), is commonly used to
select a ``sufficient" dimension for models based on a delay
reconstruction; here ``sufficient" means large enough to resolve
the dynamics without self-intersection of likely solutions in the
(projected) perfect model . Let $d_{ij}(m)$ denote the distance
between two state vectors ${\bf s}_i$ and ${\bf s}_j$ in a
reconstructed state space of dimension $m$.  This method relies on
identifying the fraction of false near neighbours, that is pairs
of points whose separations $d_{ij}(m)$ and $d_{ij}(m+1)$ satisfy
$\Delta({\bf s}_i,{\bf s}_j,m) = d_{ij}(m+1)/d_{ij}(m) > \gamma$.
A shortcoming of that approach to the detection of these false
near neighbours \cite{kennel92} is the arbitrary threshold
$\gamma$ which determines which pairs are ``false''. In a state
space where the dimension $m$ is sufficient to remove all
self-intersections $\Delta({\bf s}_i,{\bf s}_j,m)$ will be
dependent on the structure of the dynamics in the neighbourhood of
${\bf s}_i$ and ${\bf s}_j$ and thus is {\it expected} to vary
throughout the model-state space if ${\bf s}_i$ varies.

An alternative method for identifying a sufficient dimension $m$ is to
make local linear predictions using different values of $m$.
If self-intersections exist, the fraction of inconsistent points will scale
as $f_{inc} = r^{m - D_1}$, otherwise $f_{inc}$ should drop to zero for
a sufficiently small value of $r$.
In any case $\gamma$ should be allowed to vary with $x$; within
the local linear model, for example, it can vary with $sigma_1$.

\subsection{Rulkov circuit equations}

\begin{figure}
\centerline{\psfig{file=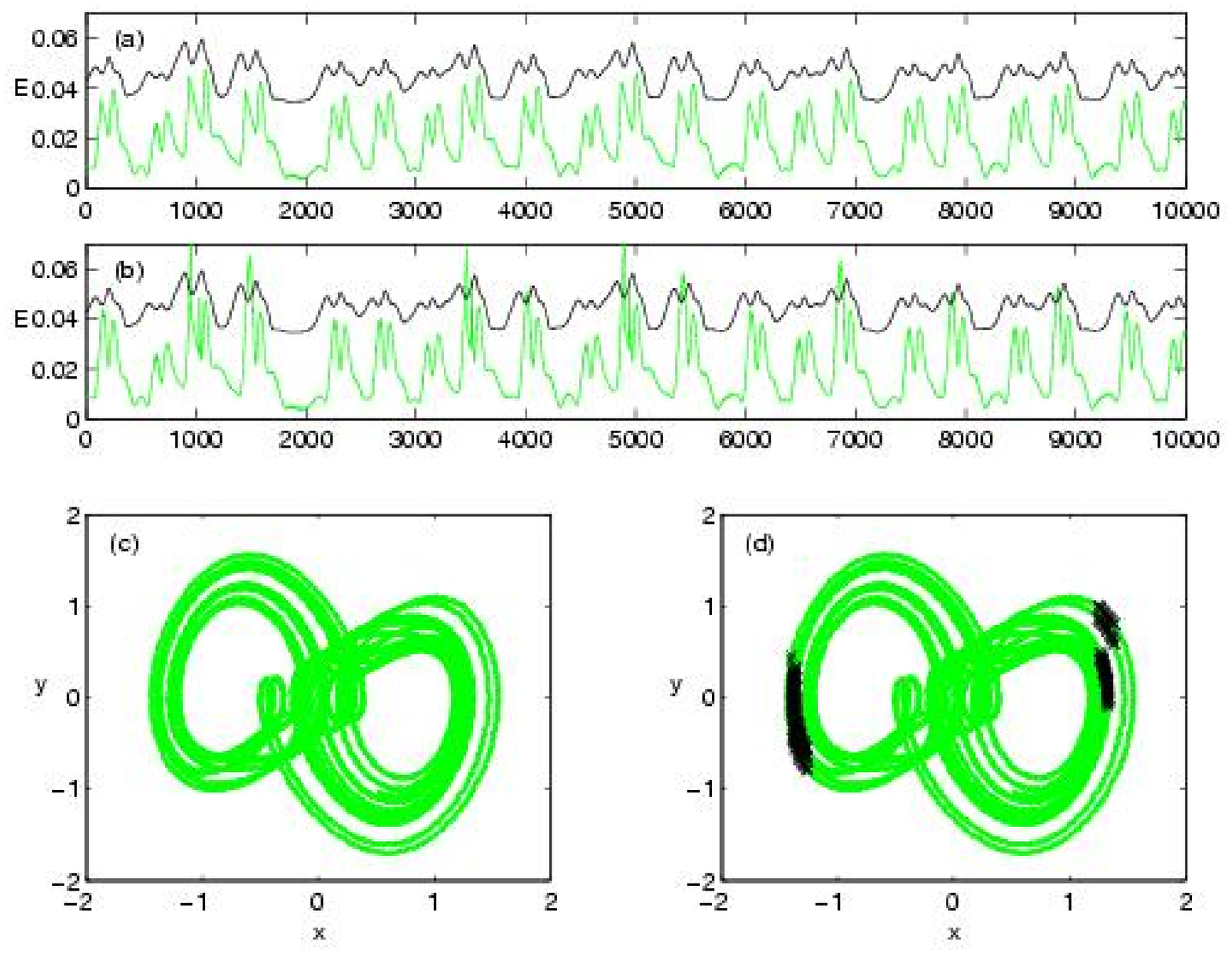,width=\linewidth}}
\caption{Consistency analysis of the Rulkov circuit.
The upper panels illustrate the prediction errors
for the perfect model (a) imperfect model (b).
The grey line is the prediction error and the black line is the consistency
envelope. Points above the envelope are inconsistent.
The lower panels are 2D projections of the delay reconstruction showing
both consistent points (grey dots) and inconsistent points ($\times$)
for perfect (c) and imperfect model (d).}
\label{f:abccon}
\end{figure}

A second mathematical system will be used to illustrate how CND
can identify state space dependent model error, in this case due
to incorrect parameter values.  The system is
the set of equations defining Rulkov's circuit
\cite{rulkov94,timmer00} are 
\bea
{\dot x} &=& y, \nonumber \\
{\dot y} &=& -x - \delta y + z, \nonumber \\
{\dot z} &=& \gamma \left[ \alpha f(x(t)) - z \right] - \sigma y,
\label{e:circuit}
\eea
where $x(t)$ is the voltage across the capacitor $C$,
$y(t) = \sqrt{\frac{L}{C}} i(t)$ with $i(t)$ the current through the
inductor $L$, and $z(t)$ is the voltage across the the capacitor $C'$.
Time has been scaled by $\frac{1}{\sqrt{LC}}$.
The parameters of this system have the following dependence on the physical
values of the circuit elements
\be
\gamma = \frac{\sqrt{LC}}{RC'}, \ \delta = r \sqrt{\frac{C}{L}},
\ \sigma = \frac{C}{C'}.
\ee

The function  $f(x)$ is \be f(x) = \left\{
\begin{array}{ll}
        0.528, & \mbox{$x \leq -x_a$}\\
        x(1-x^2), & \mbox{$-x_a < x < x_a$}\\
        -0.528, & \mbox{$x \geq x_a$} \end{array} \right.,
\label{e:fx}
\ee
and the control parameter $\alpha$ characterises the gain of the nonlinear
amplifier around $x = 0$.
The parameters of the circuit
correspond to the following values for the coefficients in the
differential equations (\ref{e:circuit}): $\gamma = 0.294$, $\sigma = 1.52$,
$\delta = 0.534$, $\alpha = 15.6$ and $x_a = 1.2$.

The model is the same set of equations but with $x_a = 1.4$,
thus the model is structurally correct, has one parameter in
error; all other parameters are exactly the same as the system.

CND is now used to map out regions of inconsistent points in the
state space.  The observations consist of three dimensional time
series of $x$, $y$ and $z$ variables, forming a state vector ${\bf x} = (x,y,z)$, 
contaminated with additive
measurement noise independently and uniformly distributed on
$[-\epsilon,\epsilon]$; $\epsilon = 0.01$. 
Numerical integration of (\ref{e:circuit}) using a fourth order Runge-Kutta 
method \cite{press92} with a fixed integration step yields a discrete-time map. 
Given an observed initial condition ${\bf x}_0 = {\bf x}(t_0)$ 
and an initial uncertainty 
$\gbf{\epsilon}(t_0)$ at time $t_0$, 
the uncertainty after an arbitrary time $\tau$ is  
\be
\gbf{\epsilon}(t_0 + \tau)
= \gbf{{\cal M}}({\bf x}_0, \tau) \gbf{\epsilon}(t_0),
\label{e:uptp}
\ee
where the matrix ${\gbf{\cal M}}({\bf x}_0, \tau)$
is the {\it linear propagator} defined by 
\be
\gbf{{\cal M}}({\bf x}_0, \tau)
= \exp \left[ \int_{t_0}^{t_0 + \tau}
\gbf{\cal J}{\bf (}{\bf x}(t){\bf )} dt \right],
\label{e:lup}
\ee
and $\gbf{\cal J}{\bf (}{\bf x}(t){\bf )}$ is the Jacobian of the flow given 
by (\ref{e:circuit}). 
Taking discrete steps of size $\tau = 0.01$, the model was then used
to provide one step ahead predictions of the 
values of the three-dimensional state vector ${\bf x} = (x,y,z)$.

CND analysis of a model with $x_a = 1.4$ can be contrasted with
that of a perfect model (that is, $x_a = 1.2$) to demonstrate how
the approach identifies regions of model-state space with
systematic errors due to parameter uncertainty.

The first singular value of the linear propagator 
${\gbf{\cal M}}({\bf x}_0, \tau)$
provides a consistency bound (\ref{e:con}) for each prediction
(Fig. \ref{f:abccon}). As expected, the perfect model with $x_a =
1.2$ yields no inconsistent points (Fig. \ref{f:abccon}a and
\ref{f:abccon}c) whereas the model with $x_a = 1.4$ gives rise to
occasional inconsistent predictions (Fig. \ref{f:abccon}b and
\ref{f:abccon}d), the location of which correctly diagnoses the
(known) model imperfection. In this way, a CND analysis can be
used to identify ``synoptic patterns" where models tend to fail.
This identification does not replace the need for insight and
physical understanding to determine how to improve the model,
rather CND merely identifies where the model is most vulnerable;
and it can be applied to the largest of numerical simulation
models.

\section{Forecasting physical systems}
\label{s:physical}

In this section CND is applied to three physical systems; in each
case the performance of previously published models are
contrasted. While the state space reconstructions appear adequate
to resolve the dynamics, these examples demonstrate that other
problems arise which are model dependent. The local linear is
inaccurate when nonlinear terms dominate in the local
neighbourhood to collect sufficient near neighbours to estimate
the model parameters.  This problem often arises when some regions
of the state space are sparsely sampled, thereby requiring a large
neighbourhood. The RBF model provides a global approximation to
the dynamics, but is likely to yield a good fit of regions of
state space with a high data density while neglecting those which
are poorly sampled \cite{smith92}. CND provides a means of
identifying and addressing all these problems and suggests a
method for selectively using different models with complementary
strengths and weaknesses.

\subsection{Power output from NH$_3$ laser}

\begin{figure}
\centerline{\psfig{file=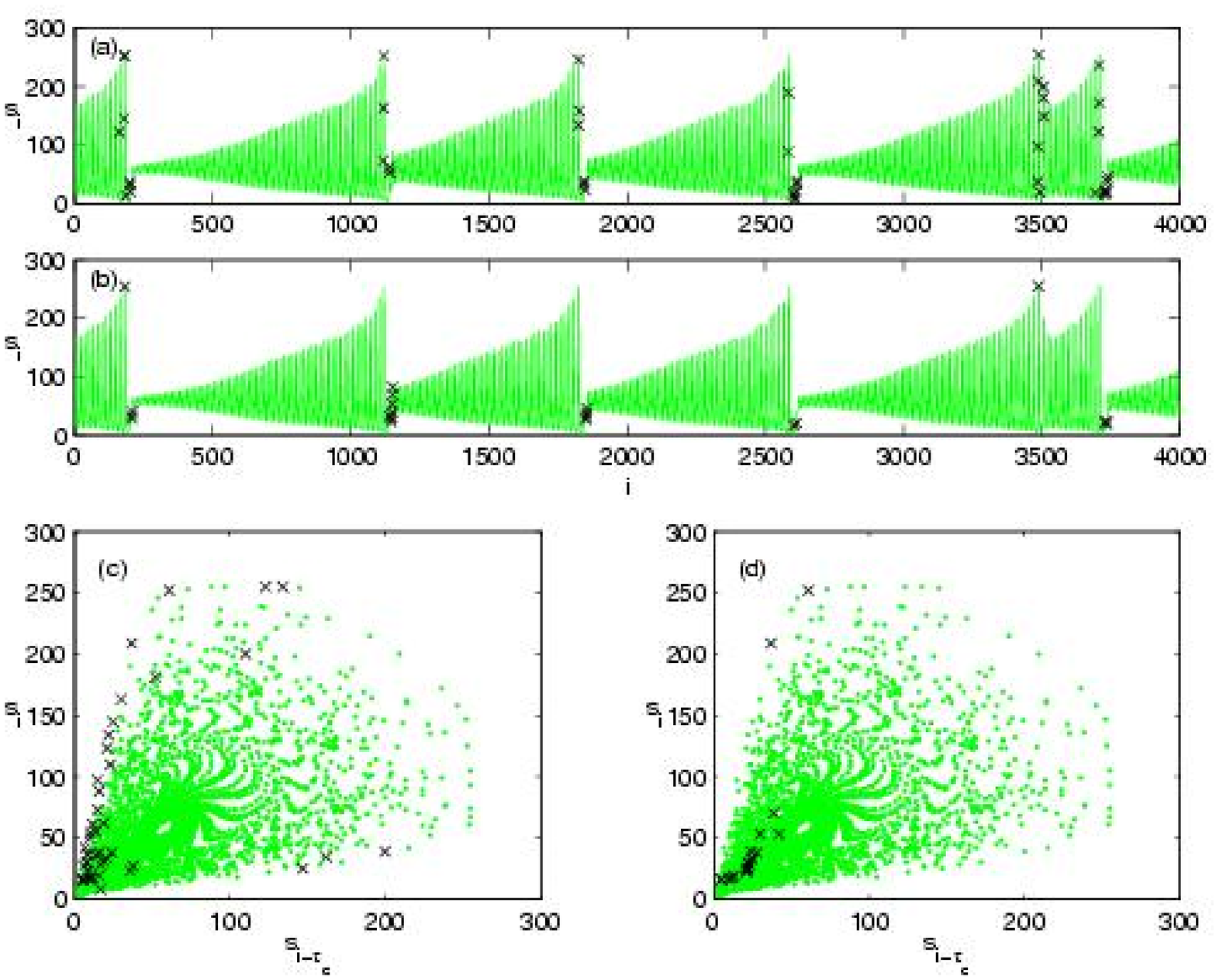,width=\linewidth}}
\caption{Predictions of the NH$_3$ laser data.
The upper panels illustrate the inconsistent predictions
($\times$) for models RBF$_1$ (a) and LL$_{1a}$ (b).
The lower panels are 2D projections of the delay reconstruction showing
both consistent points (grey dots) and inconsistent points ($\times$)
for RBF$_1$ (c) and LL$_{1a}$ (d).}
\label{f:sfcon}
\end{figure}

Fig. \ref{f:sfcon}a shows the output power of a ${\rm NH}_3$ laser
whose dynamics are associated with the Lorenz-Haken equations
\cite{haken75}; this dataset, collected by H\"{u}ner {\it et al.}
\cite{hubner93}, has been widely studied (see Weigend and
Gershenfeld \cite{weigend93}, and references thereof). The noise
level is taken to be below\footnote{Except for that due to
saturation of the analogue to digital converter, see
Smith \cite{smith93}.} the resolution of an (8 bit) analogue to
digital converter, thus $\epsilon = 1$. Following
Smith \cite{smith93}, a Local Linear model (LL$_{1a}$), was
employed to make one-step predictions and compared with a Radial
Basis Function model (RBF$_1$). Both models use the same delay
reconstruction parameters (see Table \ref{t:predres}). Typical
segments of the time series and images of inconsistent predictions
are plotted for both models in Fig. \ref{f:sfcon}. Note that all
the inconsistent points occur around the collapses; this may
result either from the relatively sparse sampling of this region
of model-state space or a local failure of the embedding
\cite{smith93}. A 2D projection of the 4D reconstructed
model-state space is illustrated in Fig. \ref{f:sfcon}c and
\ref{f:sfcon}d, showing the origins of the inconsistent
predictions for each model. If the reconstruction parameters, $m$
and $\tau_d$ are sufficient to resolve the dynamics, it may be
possible to improve the consistency by employing a neighbourhood
size which is suitable for the local linear
approximation \cite{smith94c}.

\begin{figure}
\centerline{\psfig{file=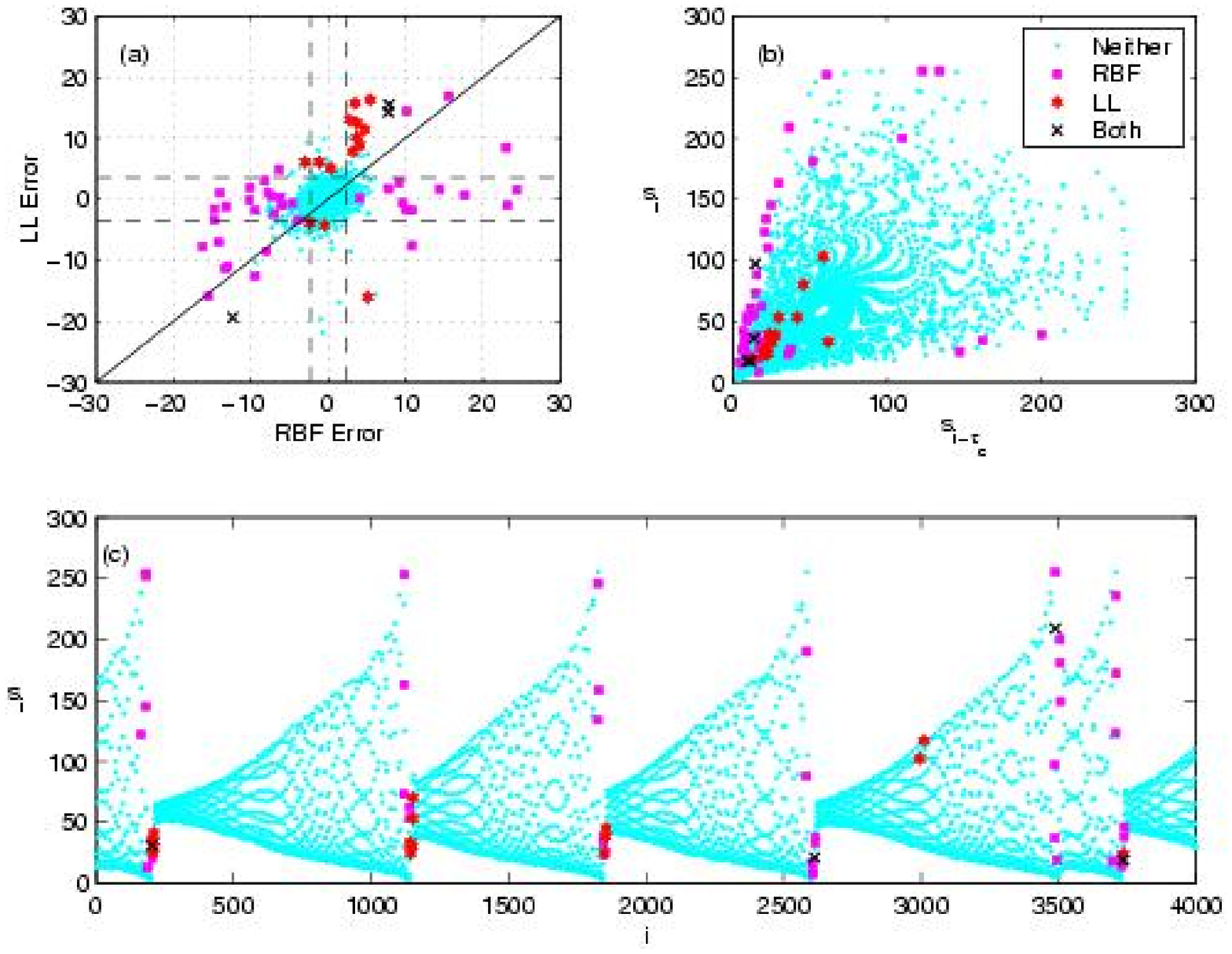,width=\linewidth}}
\caption{Consistency analysis of the NH$_3$ laser data showing
(a) prediction errors for the RBF$_1$ and LL$_{1b}$ models,
(b) 2D projections of the delay reconstruction and (c) the time series.
Markers indicate one of four
outcomes: (i) neither model is inconsistent (dot), (ii) RBF$_1$ is
inconsistent (square), (iii) LL$_{1b}$ is inconsistent (hexagon) or (iv)
both models are inconsistent (cross). Dashed lines reflect plus and minus
one standard deviation of the error distribution for each model in (a).}
\label{f:sfconrbfll}
\end{figure}

The problem of low data density around the collapses was addressed
by using a different LL model (LL$_{1b}$) with a smaller neighbourhood 
size given by $k = 8$, thereby preventing neighbours with different 
dynamics from heavily influencing the predictions in data sparse regions.  
This new LL model removes some inconsistent predictions at the beginning 
of the collapses, but adds some new inconsistent predictions at the end 
of the collapses (Fig. \ref{f:sfcon}d and stars in \ref{f:sfconrbfll}b).

The fact that some predictions are inconsistent for one model, yet
consistent for another (Fig. \ref{f:sfconrbfll}) suggests that a
hybrid predictor using both of these models selectively would
outperform any one individual model; indeed such a hybrid model
for this system is presented by Smith \cite{smith93} based on RMS
error criteria. One advantage of using a CND criteria instead is
that it focuses attention on regions of model inadequacy, even
when the one-step RMS error may be rather small. RBF predictions
with large absolute errors tend to correspond with small LL$_{1b}$
errors and vice versa (Fig. \ref{f:sfconrbfll}). While the RBF has
inconsistent predictions at the beginning of the collapses,
LL$_{1b}$ gives inconsistent predictions at the end of the
collapse, just as the intensity starts to increase again. Table
\ref{t:predres} provides a description of the models and their
prediction results and Table \ref{t:conres} gives a comparison of
the different models. In particular, while the RBF and LL$_{1b}$
models generate 1.13\% and 0.48\% inconsistent predictions
individually, they have only 0.1\% in common. Furthermore
LL$_{1b}$ provides a better complement to RBF than LL$_{1a}$ since
there are 0.15\% inconsistent predictions shared between RBF and
LL$_{1a}$.

To establish whether or not an in-sample CND analysis of the
learning set might contain information on out-of-sample forecasts,
a hybrid model was constructed. Following
Smith \cite{smith92,smith94c}, the RBF centres were used to define a
Voronoi partition\footnote{A given ${\bf s}_i$ belongs to partition
$j$, associated with centre $\gbf{\zeta}_j$, if $d_{ij} =
\min_{k=1}^{N_c} d_{ik}$ where $d_{ik} = ||{\bf s}_i -
\gbf{\zeta}_k||$.} of the state space and the fraction $p_j$ of
inconsistent predictions in each partition $j$ ($j =
1,\ldots,N_c)$ was calculated by making in-sample predictions of
the learning data set. For each base point ${\bf s}_i$ belonging
to a partition which had one or more in-sample inconsistent
predictions, a local linear model (either LL$_{1a}$ or LL$_{1b}$)
was used to generate a prediction. If the inconsistent regions of
state space are clustered in state space and the local linear
models are complementary to the RBF model, then this hybrid should
improve the prediction performance. Indeed, both the combined
models, RBF$_1$LL$_{1a}$ and RBF$_1$LL$_{1b}$ had smaller RMS
errors than their constituent models and also lower fractions of
inconsistent predictions (Table \ref{t:predres}).

\subsection{Power output from an NMR laser}

\begin{figure}
\centerline{\psfig{file=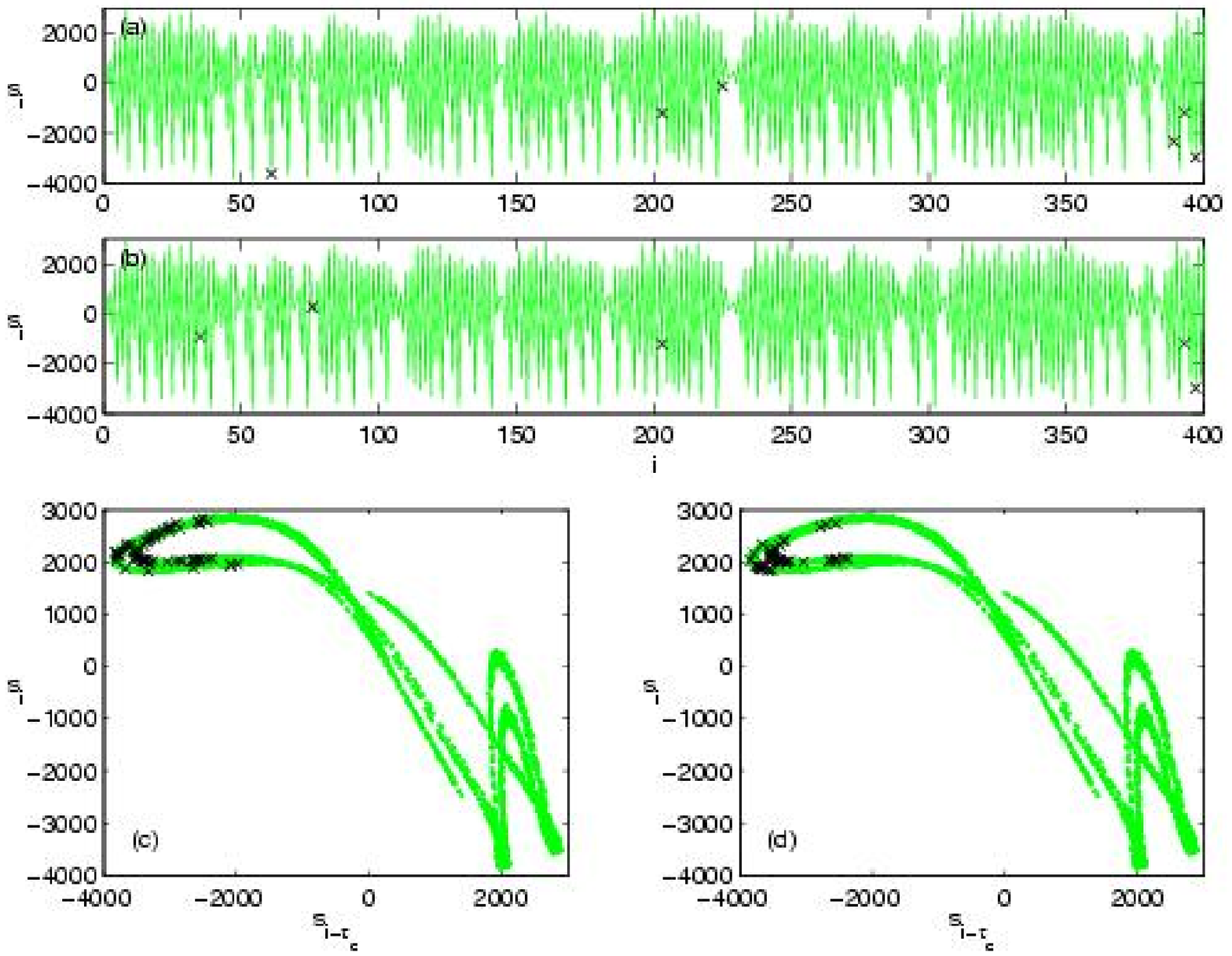,width=\linewidth}}
\caption{Predictions of the NMR laser data contrasting models
RBF$_2$ and LL$_{2a}$. The upper panels illustrate the inconsistent
predictions ($\times$) for models RBF$_2$ (a) and LL$_{2a}$ (b).
The lower panels are 2D projections of the delay reconstruction showing
both consistent points (grey dots) and inconsistent points ($\times$)
for RBF$_2$ (c) and LL$_{2a}$ (d).}
\label{f:nmrcon}
\end{figure}

Figure \ref{f:nmrcon}a illustrates a stroboscopic view of the
output power of a nuclear magnetic resonance (NMR) laser operated
at ETH Z\"{u}rich \cite{flepp91}. The lasing particles are Al
atoms in a ruby crystal, and the quality factor of the resonant
structure is modulated periodically. Output power of the laser is
reflected in the voltage across the antenna (thus allowing for
negative values). A local linear model (LL$_{2a}$) and a RBF model
(RBF$_2$) were constructed following Kantz and
Schreiber \cite{kantzbook}. Details of the models are given in
Table \ref{t:predres}. The model LL$_{2a}$ had a fixed radius
neighbourhood of size $r = 50$ (allowing the number of neighbours
to vary); this is roughly equal to twice the amplitude of the
reported measurement error amplitude, $\epsilon = 25$.
Inconsistent points are shown in Fig. \ref{f:nmrcon}a and
\ref{f:nmrcon}c for the RBF model and in Fig. \ref{f:nmrcon}b and
\ref{f:nmrcon}d for the LL$_{2a}$ model. The inconsistent
predictions typically originate from the same location of
model-state space. In this case, reconstructed vectors from
regions where a large positive observation follows a large
negative observation tended to be inconsistent.

\begin{figure}
\centerline{\psfig{file=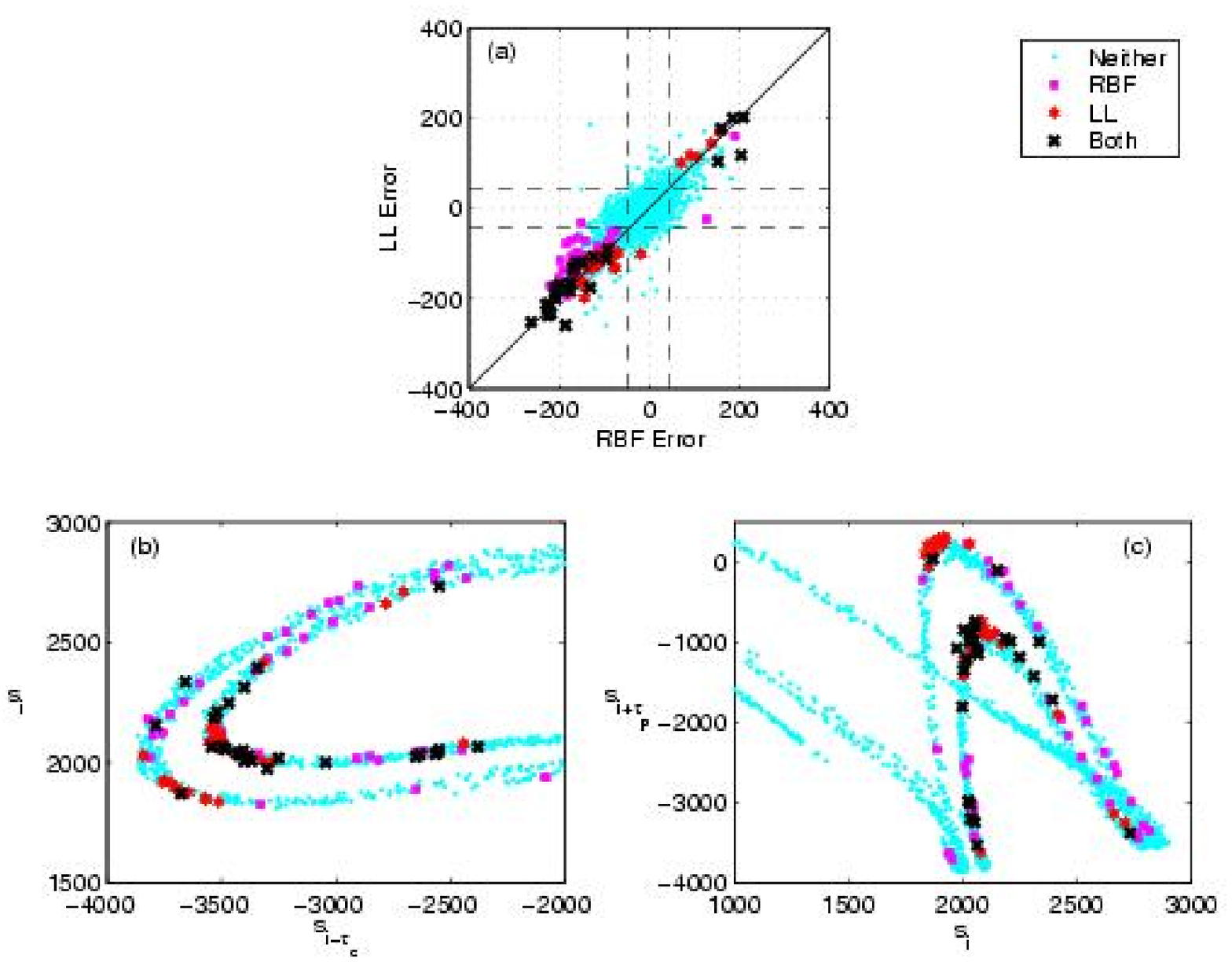,width=\linewidth}}
\caption{Consistency analysis of the NMR laser data showing
(a) prediction errors for the models RBF$_2$ and LL$_{2a}$ models,
(b) return map of $s_i$ versus $s_{i-\tau_d}$ and (c) return map of
$s_{i+\tau_p}$ versus $s_i$.
Markers indicate one of four
outcomes: (i) neither model is inconsistent (dot), (ii) RBF is
inconsistent (square), (iii) LL$_{2a}$ is inconsistent (hexagon) or (iv)
both models are inconsistent (cross).}
\label{f:nmrconrbfllw}
\end{figure}

This insight from the CND analysis suggests examining the scatter
plot of the prediction errors (Fig. \ref{f:nmrconrbfllw}a). This
figure shows that errors generated by the RBF and LL$_{2a}$ models
are highly correlated, with both making errors of similar sign and
magnitude. By zooming into the inconsistent region of the
attractor, it becomes clear that model inadequacy around the elbow
of the attractor, shown for the base of the inconsistent points
(Fig. \ref{f:nmrconrbfllw}b) and their images (Fig.
\ref{f:nmrconrbfllw}c), causes both models to fail. This effect is
most noticeable for the LL$_{2a}$ model at the elbows of the
attractor. In contrast, the ability of the LL$_{2a}$ model to
resolve dynamics in a small neighbourhood provides consistent
predictions in the upper part of the attractor (Fig.
\ref{f:nmrconrbfllw}b) whereas the RBF model is unable to resolve
the dynamics around the two leaves of the attractor.

One method for addressing the failure of the LL$_{2a}$ model to
approximate the dynamics at the elbow of the attractor (Fig.
\ref{f:nmrconrbfllw}b) is to use a local quadratic (LQ) model. A
LQ model (LQ$_{2a}$) with neighbourhood $r = 50$ did not change
the number of inconsistent predictions and increased the
prediction errors (Table \ref{t:predres}).  By decreasing the
radius to $r = 25$, a new LQ model (LQ$_{2b}$) decreased the
fraction of inconsistent predictions (from 1.4\% to 0.5\%) at the
elbow at the expense of increasing the overall prediction 
accuracy (Table \ref{t:predres}).

The benefit of using two models together may be seen from the
fractions of inconsistent predictions in the different scenarios
(Table \ref{t:conres}).
While the RBF and LQ$_{2b}$ independently have 1.65\% and 0.83\% inconsistent
predictions respectively (Table \ref{t:predres}), they only have 0.375\%
inconsistent predictions in common.  Note that the local models using
large neighbourhoods, LL$_{2a}$ and LQ$_{2a}$, offer little
compensation to the model inadequacy in the RBF model.

An alternative to changing the structure of the model is simply to
alter the local neighbourhood size; this can even lead to a
meaningful\footnote{A meaningful reduction in that the dynamics are 
more accurately reflected as opposed to, say, the reduction due to
providing the mean of a bi-modal distribution.} reduction of the
RMS error. This is illustrated in the next paragraph by taking the
inconsistent points of model (LL$_{2a}$) and predicting them with
a LL model (LL$_{2b}$) that uses a fixed number of neighbours
given by $k = 12$. While LL$_{2b}$ has a slightly higher RMS error
than LL$_{2a}$, it reduces the fraction of inconsistent
predictions from 1.4 to 0.5 (Table \ref{t:predres}).

\begin{figure}
\centerline{\psfig{file=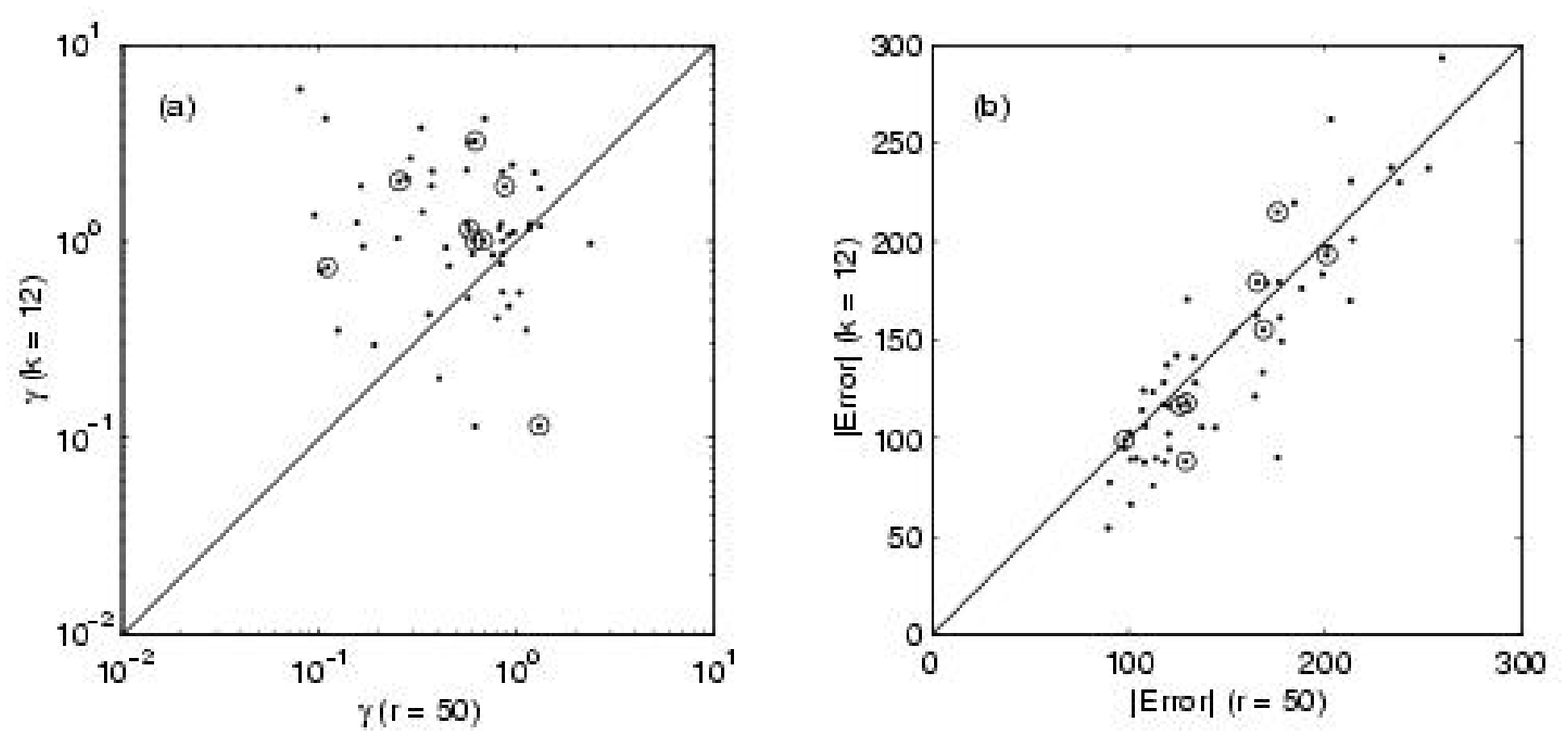,width=\linewidth}}
\caption{A comparison between models LL$_{2a}$ ($r = 50$) and
LL$_{2b}$ ($k = 12$) for predictions which are inconsistent under LL$_{2a}$:
(a) magnification factors $\gamma$ and (b) absolute prediction errors.
Circles indicate points that are also inconsistent in LL$_{2b}$.}
\label{f:nmrmag}
\end{figure}

As noted in the introduction, the appropriateness of each local
linear model can be determined by examining the expected value of
the observational noise divided by the (local) length scale at
which quadratic (higher order) terms become important. CND can
suggest where in the model-state space this fundamental ratio is
large, and in such locations one must consider local quadratic
models as above. In regions where this ratio is small, and the
length scale at which quadratic terms become important is also
small, it is advantageous to take smaller neighbourhoods thereby
improving the local linear model. In this case CND successfully
identifies such regions. Figure \ref{f:nmrmag}a shows the expected
magnification\footnote{In a delay reconstruction, the orientation
of the forecast error is known {\it a priori}. The expected
magnification, $\gamma$, is the stretching in this direction (it
need not be the first singular value); if the data is noise-free
the other directions correspond to a rotation. In practice, there
is uncertainty in all components; contrasting $\sigma_1$ and
$\gamma$ provides information about noise reduction that will be
considered elsewhere.} in a LL$_{2b}$ model plotted against the
that of the LL$_{2a}$ model for only those points where the
LL$_{2a}$ model is not consistent. The fact that the points
generally lie above the diagonal indicates that the LL$_{2b}$
model is (locally) more sensitive to uncertainty than the
LL$_{1a}$ model which averages effects over a larger area. Not
only does this increased sensitivity correctly reduce
inconsistency, it also leads to better (local) predictions, as
indicated by Fig. \ref{f:nmrmag}b, the corresponding scatter
diagram of observed prediction error. For these points the
LL$_{2b}$ model has a lower median error, lower RMS error, and
predicts 66\% of the points more accurately.

\subsection{Temperature data from a Fluid Annulus}
\label{sec:annulus}

\begin{figure}
\centerline{\psfig{file=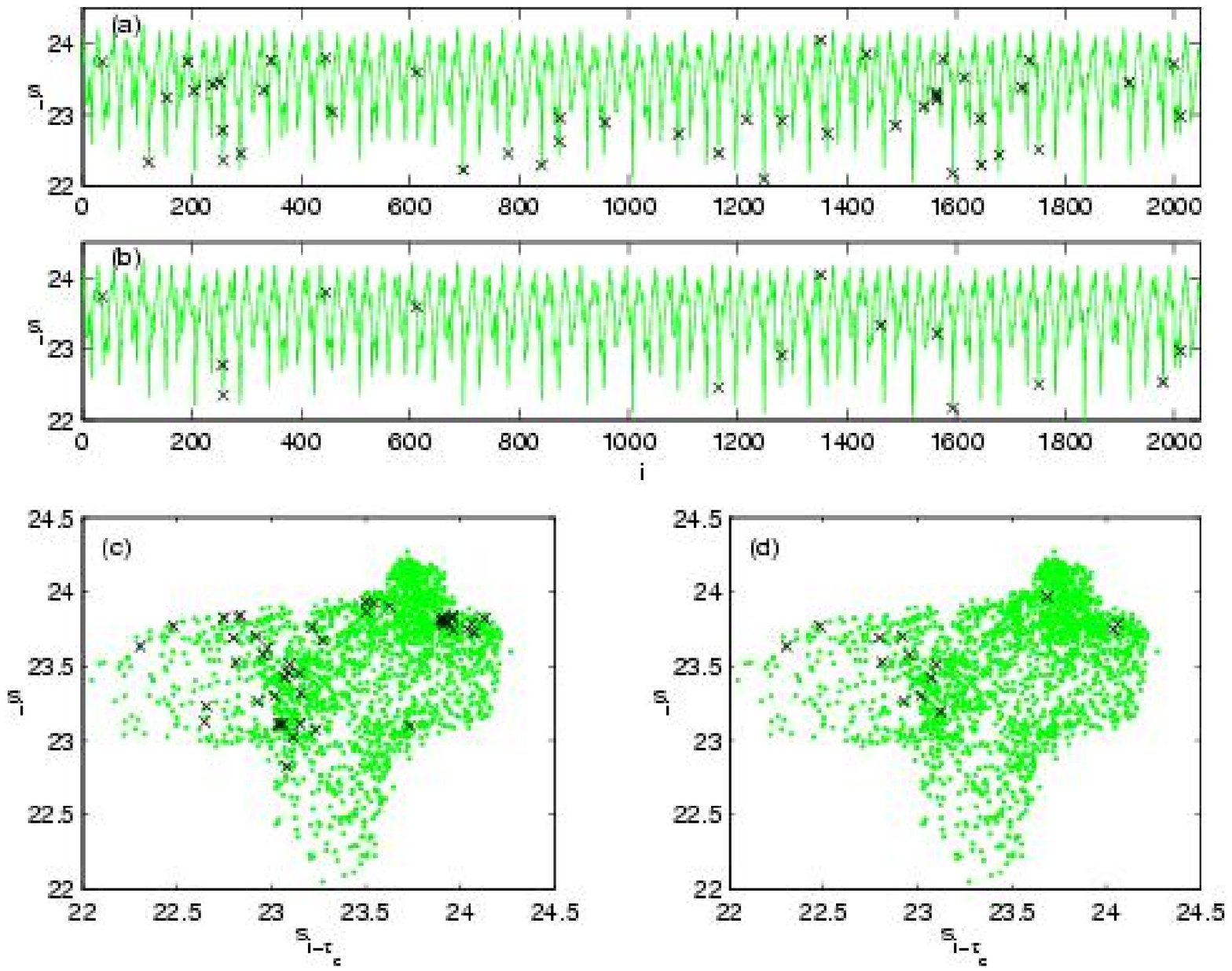,width=\linewidth}}
\caption{Predictions of the Annulus data contrasting models
RBF$_3$ and LL$_3$. The upper panels illustrate the inconsistent
predictions ($\times$) for models RBF$_3$ (a) and LL$_3$ (b).
The lower panels are 2D projections of the delay reconstruction showing
both consistent points (grey dots) and inconsistent points ($\times$)
for RBF$_3$ (c) and LL$_3$ (d).}
\label{f:anncon}
\end{figure}

Figure \ref{f:anncon}a shows a time series from a temperature
probe in a rotating annulus of fluid \cite{read92,hide58}. A
classic experiment in geophysics which Lorenz \cite{lorenz63}
cited as physical motivation for deterministic aperiodic flow, the
annulus consists of thermally conducting side walls and insulating
boundaries on the top and bottom. A temperature difference is
maintained between the inner and outer side walls, providing an
infinite dimensional simulation of the mid-latitude circulation of
the Earth's atmosphere. Following \cite{smith92}, one step ahead
predictions were made using a local linear model (LL$_3$) and a
RBF model (RBF$_3$). See Table \ref{t:predres} for details of the
models. The data was assumed to have uniformly distributed noise
of amplitude $\epsilon = 0.1$. The inconsistent points (Fig.
\ref{f:anncon}) reveal that the LL model is extremely good
($f_{inc} = 0.68\%$) whereas the RBF model ($f_{inc} = 2.25\%$)
yields a much larger number of inconsistent predictions.

\begin{figure}
\centerline{\psfig{file=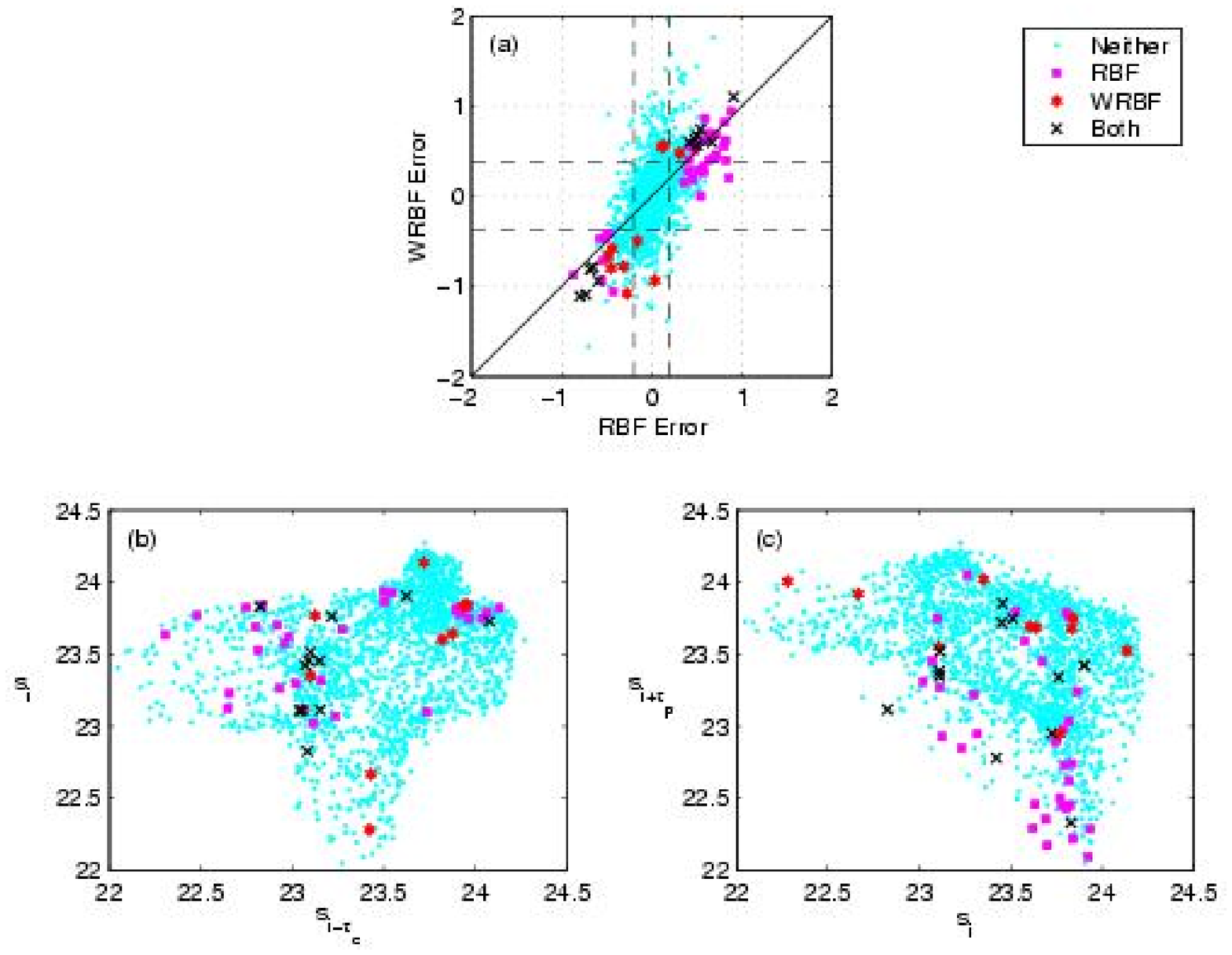,width=\linewidth}}
\caption{Consistency analysis of the Annulus data contrasting models
RBF and WRBF:
(a) prediction errors for the models RBF$_3$ and WRBF,
(b) return map of $s_i$ versus $s_{i-\tau_d}$ and (c) return map of
$s_{i+\tau_p}$ versus $s_i$.
Markers indicate one of four
outcomes: (i) neither model is inconsistent (dot), (ii) RBF$_3$ is
inconsistent (square), (iii) WRBF is inconsistent (star) or (iv)
both models are inconsistent (cross).}
\label{f:annconrbfrbfw}
\end{figure}

One approach for improving the consistency of the RBF model is to
force it to provide a better fit to the dynamics in regions of
state space which were inconsistent in the learning data set. The
RBF centres were used to form a Voronoi partition of the state
space and the fraction $p_j$ of inconsistent predictions in each
partition $j$ ($j = 1,\ldots,N_c)$ was calculated using in-sample
predictions of points in the learning data set. Following Smith
\cite{smith92} which re-weighted partitions to get a more uniform
error in the model-state space,  CND analysis provides the
information required to construct an RBF model which better
approximates the more inconsistent partitions (those with large
values of $p_j$). This is done by providing weights ${\bf w} = \{
w_i \}_{i=1}^{N_l}$ such that $w_i = 1 + \alpha p_{n(i)}$ where
$n(i)$ is the index of the partition containing ${\bf s}_i$, and
computing new parameters ${\bf a} = ({\bf w} \gbf{\Phi})^\dag
({\bf w} {\bf b})$. This new weighted RBF (hereafter, WRBF) model
will have a higher RMS error since, by definition, the RMS error
is minimised when $w_i = 1$ for all $i$. While the prediction
errors will increase in the partitions with small $p_j$, the WRBF
can be used to reduce the overall number of inconsistent
predictions. The value of $\alpha$ controls the balance between
increasing the RMS error and improving the consistency. Using
$\alpha = 128$, the fraction of inconsistent predictions was
decreased, from $2.25\%$ for the RBF model to $1.07\%$ for the
WRBF model, whereas the RMS (normalised by the standard deviation
of the data) increased from 0.47 to 0.89. Many of the large RBF
errors were improved by the WRBF (Fig. \ref{f:annconrbfrbfw}a)
yielding less inconsistent predictions (Fig.
\ref{f:annconrbfrbfw}b and \ref{f:annconrbfrbfw}c). The fraction
of inconsistent predictions shared by the RBF and WRBF is only
0.586\% (Table \ref{t:conres}). This small fraction is equal to
the fraction of shared inconsistent predictions for the RBF and
the LL.  While it is to be expected that the RBF and LL would
yield different inconsistent predictions because of the disparity
between their model structures, this result suggests that the
weights provide a means of obtaining a complementary version of a
given RBF model.

A hybrid model RBFWRBF was constructed using the RBF to make predictions
in all partitions except those where there were one or more inconsistent
predictions in the learning set.  In this case the inconsistent predictions
are not clustered and so the hybrid gives a medium level of performance,
having a lower RMS error than the WRBF but a higher RMS error than the RBF.

\section{Discussion}
\label{s:discussion}

A new test for Consistent Nonlinear Dynamics (CND) has been
introduced and illustrated on a number of examples. The key aim of
CND is to quantify the consistency between a model's dynamics and
the observations locally throughout the model-state space. Regions
of systematic failure indicate states of the system where the
model needs improvement, regardless of whether the absolute value
of the errors in that region are large or small. Similarly,
regions of large errors which are consistent with the model
dynamics and level of observational uncertainty should not be
counted against the model {\it a priori}. In such regions the
observations are consistent with the model forecasts, and the
model can be accepted. Of course, this acceptance is provisional,
the model may well fail to remain consistent when the noise level
is reduced, or another (locally) consistent model may be found to
have better error statistics in this region and longer shadowing
times. In any event, lower one-step forecast errors {\it per se}
need not indicate a better model. The errors of the best first
guess forecast are simply beside the point: the simplest model
consistent with the data should admit trajectories consistent with
the observations.

Where the model has been consistent, prognostic assessment of the
likely accuracy of predictions can be made in real-time, while in
regions of the model state space where the model tends to be
inconsistent model based estimates of predictability (whether
analytic, or made with ensemble forecasts \cite{palmer00}) should
not be relied upon; in such regions historical errors may prove 
to be of value \cite{smith92}.

Other insights from the application of CND include:

(i) The distribution of prediction errors from a nonlinear
model is expected to show correlation in state space and thus show
significant residual predictability \cite{smith94c,mcsharry99a}. Equation
(\ref{e:enoisete}) shows that this is to be expected even with a
perfect nonlinear model: the local residuals need not be
symmetrically distributed about the global mean error
\cite{mcsharry99a,dphil99}.

(ii) The original method of false nearest neighbours
\cite{kennel92} can be strengthened, replacing the arbitrary
global threshold used to define ``false'' by adopting a local
threshold based on the internal consistency of the dynamics.

(iii) Noise reduction strategies often use a variational approach
\cite{palmer00}; the interpretation of a variational data
assimilation scheme assumes the existence of a consistent model
trajectory. CND can verify this assumption.

(iv) The `optimal linear predictor' is often inconsistent with the
observations in a systematic manner; CND will detect this.

(v) All models of physical systems contain structural error; no
single {\it best} model need exist (for discussion, see
\cite{smith00,smithecmwf2003,rapp99}). Rather
than trying to obtain a single optimal model, it may be prove 
effective to consider ensembles over models with different model
structures. Ensembles over different initial conditions are often
utilised to account for observational uncertainty
\cite{smith97fermi,smith00,palmer00}. Results from this paper
suggest a complementary method which accounts for structural
uncertainty by employing ensembles over model structures (see
\cite{dphil99,smithecmwf2003}).

(vi) Systematic changes in the frequency or location of
inconsistent points may indicate non-stationarity in the
underlying process \cite{smith92}.

(vii) Given a data stream, selective refinement of points corresponding to
inconsistent regions is expected to yield a better data base for
local models than uniform sampling, at least in the noise-free case
for local polynomial models \cite{smith94c,dphil99,kwasnioks04}.

\begin{table}
\begin{center}
\caption{Description of the models used for the three physical systems.
Reconstruction parameters are dimension $m$, time delay $\tau_d$,
and dataset sizes used for learning $N_l$, and testing $N_t$ (in 1000s).
$N_c$ is the number of RBF centres and $\phi(r)$ gives the structure of the
RBF.  The LL neighbourhood is defined by
the number of neighbours $k$ or the radius $r$.
Error is the RMS forecast error
normalised by the standard deviation of the data and $\%f_{inc}$ is the
percentage of inconsistent predictions.}
\begin{tabular}{lllllllllllll}
Key  & System  &$m$  &$\tau_d$  &$\tau_p$  &$N_l$  &$N_t$
&$N_c$  &$\phi(r)$  &$k$  &$r$  &Error  &$\%f_{inc}$  \\
\hline
RBF$_1$  &NH$_3$  &4  &2  &1 &21  &4  &128 &$r^3$  & & &0.0514  &1.13 \\

LL$_{1a}$ &NH$_3$  &4  &2  &1 &21  &4  &  &  &32  &    &0.0381  &0.53 \\

LL$_{1b}$ &NH$_3$  &4  &2  &1 &21  &4  & & &8 &    &0.0797  &0.48 \\

RBF$_1$LL$_{1a}$ &NH$_3$  &4  &2  &1 &21  &4  &128  &$r^3$  &32 &
&0.0376  &0.30 \\

RBF$_1$LL$_{1b}$ &NH$_3$  &4  &2  &1 &21  &4  &128  &$r^3$  &8  &
&0.0451  &0.27 \\

\hline
RBF$_2$  &NMR     &3  &1  &1 &30  &4  &128 &$r^3$ & & &0.0207  &1.65 \\

LL$_{2a}$ &NMR    &3  &1  &1 &30  &4  & & & &50     &0.0189  &1.40 \\

LL$_{2b}$ &NMR    &3  &1  &1 &30  &4  & & &12 &     &0.0197  &0.50 \\

LQ$_{2a}$ &NMR     &3  &1  &1  &30  &4  & & & &50     &0.0210  &1.40 \\

LQ$_{2b}$ &NMR     &3  &1  &1  &30  &4  & & & &25     &0.0243  &0.83 \\
\hline
RBF$_3$  &Annulus &5  &4  &18 &2  &4  &128 &$e^{-r^2/2\sigma_2}$ & &
&0.4764  &2.25 \\

LL$_3$   &Annulus &5  &4  &18 &2  &2  & & &32 &      &0.4207  &0.68 \\

WRBF  &Annulus &5  &4  &18  &2  &2  &128  &$e^{-r^2/2\sigma_2}$  & &
 &0.8989  &1.07 \\
RBFWRBF  &Annulus &5  &4  &18  &2  &2  &128  &$e^{-r^2/2\sigma_2}$  & &
 &0.5402  &1.27 \\
\end{tabular}
\label{t:predres}
\end{center}
\end{table}

\begin{table}
\begin{center}
\caption{Comparison between two models A and B.  Percentage
of predictions falling into each of four scenarios: (i) Neither A nor B are
inconsistent, (ii) only A is inconsistent, (iii) only B is inconsistent and
(iv) both A and B are inconsistent.}
\begin{tabular}{lllllll}
System   &Model A  &Model B    &(i)  &(ii) &(iii)  &(iv) \\
\hline
NH$_3$   &RBF$_1$  &LL$_{1a}$  &98.500 &0.975 &0.375 &0.150 \\
NH$_3$   &RBF$_1$  &LL$_{1b}$  &98.500 &1.025 &0.375 &0.100 \\
\hline
NMR      &RBF$_2$  &LL$_{2a}$  &97.675 &0.925 &0.700 &0.700 \\
NMR      &RBF$_2$  &LL$_{2b}$  &98.000 &1.500 &0.375 &0.125 \\
NMR      &RBF$_2$  &LQ$_{2a}$  &97.600 &1.000 &0.775 &0.625 \\
NMR      &RBF$_2$  &LQ$_{2b}$  &97.925 &1.250 &0.450 &0.375 \\
NMR      &LL$_{2a}$ &LL$_{2b}$ &98.300 &1.200 &0.300 &0.200 \\

\hline
Annulus  &RBF$_3$  &LL$_{3}$   &97.656 &1.660 &0.098 &0.586 \\
Annulus  &RBF$_3$  &WRBF       &97.266 &1.660 &0.489 &0.586 \\
\end{tabular}
\label{t:conres}
\end{center}
\end{table}

\section{Conclusion}
\label{s:conclusion} A new approach for finding the limitations of
dynamical models has been proposed and illustrated. By examining
the consistency of the local model dynamics with the observed
system dynamics, the consistent nonlinear dynamics (CND) approach
can identify regions of the model-state space where the model is
systematically inconsistent. Analysis of two theoretical systems
and three physical systems illustrates that this novel consistency
test contains a wealth of information about a model's ability to
approximate the observed dynamics in general, and delay
reconstructions in particular. If the delay reconstruction does
not yield an embedding, then all models will be inconsistent in
regions where the embedding fails. When an embedding does exist,
distinct model structures may be preferred in certain regions; CND
identifies these explicitly. Each analysis yields a direct
assessment of individual predictions, taking the local properties
of the model structure into account, thereby avoiding biases
arising from model sensitivity.

Uncovering the failure or success of candidate model structures
provides a useful discriminator for contrasting different models.
The ultimate aim here is to find better models, and a better way
to define "better" in this context. Locally consistent models are
expected to allow longer shadowing trajectories; a comparison
along these lines will be presented elsewhere.

By adopting CND as a goal, one aims to get the simplest models
consistent with the data, but none simpler. RMS skill scores can
prove a distraction in this quest.
By addressing the interplay between the nonlinear model structure
and observational uncertainty in the measurement process, the CND
approach opens many avenues for both the evaluation and the 
application of nonlinear models.

\section*{Acknowledgements}
This work was supported by EC grant ERBFMBICT950203, EPSRC grant GR/N02641
and ONR grant N00014-99-1-0056.  We are happy to acknowledge useful discussion 
with Kevin Judd and David Orrell. 


\begin{thebibliography}{10}

\bibitem{feynman}
R.~P. Feynman.
\newblock {\em The Character of Physical Law}.
\newblock Penguin Books, London, 1992.

\bibitem{chatfield89}
C.~Chatfield.
\newblock {\em The Analysis of Time Series}.
\newblock Chapman and Hall, London, New York, 4th edition, 1989.

\bibitem{mcsharry99a}
P.~E. McSharry and L.~A. Smith.
\newblock Better nonlinear models from noisy data: Attractors with maximum
  likelihood.
\newblock {\em Phys. Rev. Lett.}, 83(21):4285--4288, 1999.

\bibitem{smith97fermi}
L.~A. Smith.
\newblock The maintenance of uncertainty.
\newblock In G.~Cini, editor, {\em Nonlinearity in Geophysics and
  Astrophysics}, volume CXXXIII of {\em International School of Physics
  ``{Enrico Fermi}''}, pages 177--246, Bologna, Italy, 1997. Societ\`{a}
  Italiana di Fisica.

\bibitem{smith00}
L.~A. Smith.
\newblock Disentangling uncertainty and error: On the predictability of
  nonlinear systems.
\newblock In A.~I. Mees, editor, {\em Nonlinear Dynamics and Statistics}, pages
  31--64, Boston, 2000. Birkhauser.

\bibitem{grebogi90}
C.~Grebogi, S.~M. Hammel, J.~A. Yorke, and T.~Sauer.
\newblock Shadowing of physical trajectories in chaotic dynamics: Containment
  and refinement.
\newblock {\em Phys. Rev. Lett.}, 65:1527--1530, 1990.

\bibitem{palmer00}
T.~N. Palmer.
\newblock Predicting uncertainty in forecasts of weather and climate.
\newblock {\em Rep. Prog. Phys.}, 63:71--116, 2000.

\bibitem{orrell01}
D.~Orrell, L.~A. Smith, J.~Barkmeijer, and T.~N. Palmer.
\newblock Model error in weather forecasting.
\newblock {\em Nonlinear Processes in Geophysics}, 8(6):357--371, November
  2001.

\bibitem{kantzbook}
H.~Kantz and T.~Schreiber.
\newblock {\em Nonlinear Time Series Analysis}.
\newblock Cambridge University Press, Cambridge, 1997.

\bibitem{farmers87}
J.~D. Farmer and J.~J. Sidorowich.
\newblock Predicting chaotic time series.
\newblock {\em Phys. Rev. Lett.}, 59(8):845--848, 1987.

\bibitem{priestly81}
M.~B. Priestly.
\newblock {\em Spectral Analysis and Time Series}.
\newblock Academic Press, London, 1981.

\bibitem{brooml88}
D.~S. Broomhead and D.~Lowe.
\newblock Multivariable functional interpolation and adaptive networks.
\newblock {\em J. Complex Systems}, 2:321--355, 1988.

\bibitem{casdagli89}
M.~Casdagli.
\newblock Nonlinear prediction of chaotic time series.
\newblock {\em Physica D}, 35:335--356, 1989.

\bibitem{smith92}
L.~A. Smith.
\newblock Identification and prediction of low-dimensional dynamics.
\newblock {\em Physica D}, 58:50--76, 1992.

\bibitem{kennel92}
M.~B. Kennel, R.~Brown, and H.~D.~I. Abarbanel.
\newblock Determining embedding dimension for the phase-space reconstruction
  using a geometrical construction.
\newblock {\em Phys. Rev. A}, 45(6):3403--3411, 1992.

\bibitem{tong90}
H.~Tong.
\newblock {\em Non-Linear Time Series Analysis}.
\newblock Oxford Univ. Press, Oxford, 1990.

\bibitem{kostelich90}
E.~J. Kostelich and J.~A. Yorke.
\newblock Noise-reduction -- finding the simplest dynamic system consistent
  with the data.
\newblock {\em Physica D}, 41(2):183--196, 1990.

\bibitem{judds04}
K.~Judd and L.~A. Smith.
\newblock Indistinguishable states {II}: imperfect model scenario.
\newblock {\em Physica D}, 2004.
\newblock in review.

\bibitem{beven01}
K.~J. Beven.
\newblock Equifinality, data assimilation, and uncertainty estimation in
  mechanistic modelling of complex environmental systems.
\newblock {\em J. Hydrology}, 249:11--29, 2001.

\bibitem{beven02}
K.~J. Beven.
\newblock Towards a coherent philosophy for modelling the environment.
\newblock {\em Proc. Roy. Soc. A}, 458:2465--2484, 2002.

\bibitem{orrell}
D.~Orrell.
\newblock {\em Modelling nonlinear dynamical systems: chaos, uncertainty, and
  error}.
\newblock PhD thesis, University of Oxford, 2001.

\bibitem{schroer98}
C.~G. Schroer, T.~Sauer, E.~Ott, and J.~A. Yorke.
\newblock Predicting chaos most of the time from embeddings with
  self-intersections.
\newblock {\em Phys. Rev. Lett.}, 80(7):1410--1413, 1998.

\bibitem{dphil99}
P.~E. McSharry.
\newblock {\em Innovations in Consistent Nonlinear Deterministic Prediction}.
\newblock PhD thesis, University of Oxford, 1999.

\bibitem{takens81}
F.~Takens.
\newblock Detecting strange attractors in fluid turbulence.
\newblock In D.~Rand and L.~S. Young, editors, {\em Dynamical Systems and
  Turbulence}, volume 898, page 366, New York, 1981. Springer-Verlag.

\bibitem{sauer91}
T.~Sauer, J.~A. Yorke, and M.~Casdagli.
\newblock Embedology.
\newblock {\em J. Stats. Phys.}, 65:579--616, 1991.

\bibitem{press92}
W.~H. Press, B.~P. Flannery, S.~A. Teukolsky, and W.~T. Vetterling.
\newblock {\em {N}umerical {R}ecipes in {C}}.
\newblock CUP, Cambridge, 2nd edition, 1992.

\bibitem{ikeda79}
K.~Ikeda.
\newblock Multiple-valued stationary state and its instability of the
  transmitted light by a ring cavity system.
\newblock {\em Optics Communications}, 30(2):257--261, 1979.

\bibitem{ikeda80}
K.~Ikeda and H.~Daido.
\newblock Optical turbulence: chaotic behaviour of transmitted light from a
  ring cavity.
\newblock {\em Phys. Rev. Lett.}, 45(9):709--712, 1980.

\bibitem{hammel85}
S.~M. Hammel, C.~K. R.~T. Jones, and J.~V. Moloney.
\newblock Global dynamical behaviour of the optical field in a ring cavity.
\newblock {\em J. Opt. Soc. Am. B}, 2(4):552--564, April 1985.

\bibitem{smith94c}
L.~A. Smith.
\newblock Local optimal prediction: Exploiting strangeness and the variation of
  sensitivity to initial condition.
\newblock {\em Phil Trans R Soc Lond A}, 348(1688):371--381, 1994.

\bibitem{bollt00}
E.~M. Bollt.
\newblock Model selection, confidence, and scaling in predicting chaotic
  time-series.
\newblock {\em Int. J. Bif. Chaos}, 10(6):1407--1422, 2000.

\bibitem{kaplany79}
J.~L. Kaplan and J.~A. Yorke.
\newblock Chaotic behaviour of multidimensional difference equations.
\newblock In H.~O. Peitgen and H.~O. Walter, editors, {\em Functional
  Differential Equations and Approximation of Fixed Points}, volume 730 of {\em
  Lecture Notes in Mathematics}, page 204. Springer, Berlin, 1979.

\bibitem{rulkov94}
N.~F. Rulkov, L.S. Tsimring, and H.~D.~I. Abarbanel.
\newblock Tracking unstable orbits in chaos using dissipative feedback control.
\newblock {\em Phys. Rev. E}, 50(1):314--324, 1994.

\bibitem{timmer00}
J.~Timmer, H.~Rust, W.~Horbelt, and H.~U. Voss.
\newblock Parametric, nonparametric and parametric modelling of a chaotic
  circuit time series.
\newblock {\em Physics Letters A}, 274:123--134, 2000.

\bibitem{haken75}
H.~Haken.
\newblock Analogy between higher instabilities in fluids and lasers.
\newblock {\em Phys. Lett. A}, 53:77, 1975.

\bibitem{hubner93}
U.~H\"{u}bner, C.~O. Weiss, N.~B. Abraham, and D.~Tsang.
\newblock Lorenz-like chaos in ${\rm nh}_3$-fir lasers.
\newblock In A.~Weigend and N.~Gershenfeld, editors, {\em Time Series
  Prediction: Forecasting the Future and Understanding the Past}, volume~XV of
  {\em SFI Studies in Complexity}, pages 323--344, New York, 1993.
  Addison-Wesley.

\bibitem{weigend93}
A.~S. Weigend and N.~A. Gershenfeld.
\newblock {\em Time Series Prediction: Forecasting the Future and Understanding
  the Past}, volume~XV of {\em SFI Studies in Complexity}.
\newblock Addison-Wesley, New York, 1993.

\bibitem{smith93}
L.~A. Smith.
\newblock Does a meeting in {S}anta {F}e imply chaos?
\newblock In A.~Weigend and N.~Gershenfeld, editors, {\em Time Series
  Prediction: Forecasting the Future and Understanding the Past}, volume~XV of
  {\em SFI Studies in Complexity}, pages 323--344, New York, 1993.
  Addison-Wesley.

\bibitem{flepp91}
L.~Flepp, R.~Holzner, E.~Brun, M.~Finardi, and R.~Badii.
\newblock Model identification by periodic-orbit analysis for {NMR}-laser
  chaos.
\newblock {\em Phys. Rev. Lett.}, 67:2244--2247, 1991.

\bibitem{read92}
P.~L. Read.
\newblock Applications of singular systems to `baroclinic chaos'.
\newblock {\em Physica D}, 58:455--468, 1992.

\bibitem{hide58}
R.~Hide.
\newblock An experimental study of thermal convection in a rotating liquid.
\newblock {\em Phil. Trans. Roy. Soc. A}, 250:441--478, 1958.

\bibitem{lorenz63}
E.~N. Lorenz.
\newblock Deterministic nonperiodic flow.
\newblock {\em J. Atmos. Sci.}, 20:130--141, 1963.

\bibitem{smithecmwf2003}
L.~A. Smith.
\newblock Predictability past predictability present.
\newblock In {\em Seminar on predictability}, pages 219--242, ECMWF, Reading,
  UK, 2003.

\bibitem{rapp99}
P.~E. Rapp, T.~I. Schmah, and A.~I. Mees.
\newblock Models of knowing and the investigation of dynamical systems.
\newblock {\em Physica D}, 132:133--149, 1999.

\bibitem{kwasnioks04}
F.~Kwasniok and L.~A. Smith.
\newblock Real-time construction of optimized predictors from data streams,
  2004.
\newblock accepted.

\end{thebibliography}

\end{document}